\documentclass{article}
\usepackage[latin1]{inputenc}
\usepackage{amsmath}
\newcommand{\p}{\partial}
\newcommand{\al}{\alpha}
\newcommand{\be}{\beta}
\newcommand{\ep}{\epsilon}
\newcommand{\de}{\delta}
\newcommand{\ti}{\tilde}

\newcommand{\om}{\omega}

\usepackage{indentfirst}
\title{Non-Grassmann mechanical model of the Dirac equation}
\author{A. A. Deriglazov$^{1,}$\footnote{alexei.deriglazov@ufjf.edu.br ~
On leave of absence from Dept. Maths and Physics, Tomsk Polytechnical University, Tomsk, Russia.}, \and B. F.
Rizzuti$^{2,}$\footnote{brunorizzuti@ufam.edu.br}, \and G. P. Zamudio$^{1}$, \and P. S. Castro$^{1}$}

\date{${}^1$Dept. de Matemática, ICE, \\
Universidade Federal de Juiz de Fora, MG, Brasil \\
${}^2$ISB, Universidade Federal do Amazonas, Coari-AM, Brazil.}

\begin{document}
\maketitle \large

\begin{abstract}
We construct a new example of the spinning-particle model without Grassmann variables. The spin degrees of freedom are
described on the base of an inner anti-de Sitter space. This produces both $\Gamma^\mu$ and
$\Gamma^{\mu\nu}$\,-matrices in the course of quantization. Canonical quantization of the model implies the Dirac
equation. We present the detailed analysis of both the Lagrangian and the Hamiltonian formulations of the model and
obtain the general solution to the classical equations of motion. Comparing {\it Zitterbewegung} of the spatial
coordinate with the evolution of spin, we ask on the possibility of space-time interpretation for the inner space of
spin. We enumerate similarities between our analogous model of the Dirac equation and the two-body system subject to
confining potential which admits only the elliptic orbits of the order of de Broglie wave-length. The Dirac equation
dictates the perpendicularity of the elliptic orbits to the direction of center-of-mass motion.
\end{abstract}

\noindent

\section{Introduction}
Although the true understanding of spin is achieved in the framework of quantum electrodynamics, a lot of efforts has
been spent in attempts to construct the relativistic mechanical model of a spinning electron [2-19]. Just after the
introduction of the electron spin as a new quantum number by Pauli \cite{wp}, Uhlenbeck and Goudsmit suggested its
naive interpretation in terms of the inner angular momentum \cite{ug}. The first papers devoted the semiclassical
description of spin dynamics can be traced back up to Frenkel\footnote{See a modern revision of Frenkel´s paper in
\cite{ter}.} \cite{fr} and Thomas \cite{th}. Bargmann, Michel and Telegdi have demonstrated \cite{bmt} that the models
practically exactly reproduces the spin dynamics of polarized beams in uniform fields. However, the miracle is that the
models based on these schemes do not produce the Dirac equation through the canonical quantization. One possible
solution to the problem has been found by Berezin and Marinov \cite{bm75, GG1} in the framework of Grassmann mechanics.
The problem here is that the Grassmann mechanics represents a rather formal mathematical construction. It leads to
certain difficulties \cite{bm75, gt} in attempts to use it for the description of spin effects on the semiclassical
level, before the quantization. We also point out that there is no generalization of Grassmann mechanics on higher
spins \cite{aad9}.

Hence it would be interesting to construct the non-Grassmann model of the Dirac equation. While the problem has a long
history (see [2-7, 10-19, 23, 24, 26] and references therein), there appears to be no wholly satisfactory solution to
date. Our believe on the existence of such a kind model is based on the following well-known observation. The Dirac
spinor $\Psi$ can be used to construct the four-dimensional current vector, $\bar\Psi\Gamma^\mu\Psi$, which is
preserved for solutions to the Dirac equation, $\partial_\mu(\bar\Psi\Gamma^\mu\Psi)=0$. Hence its null-component,
$\Psi^\dagger\Psi\ge 0$, admits the probabilistic interpretation, and we expect that one-particle sector of the Dirac
equation admits description in the framework of relativistic quantum mechanics (RQM). So we can look for the
corresponding semiclassical model, which would lead to the Dirac equation in the course of canonical quantization.

However, it is well-known that adopting the RQM interpretation, we arrive at the rather strange and controversial
picture. To remind this, we use the Dirac matrices $\alpha^i$ and $\beta$, to represent the Dirac equation in the form
of the Schr\"odinger equation
\begin{eqnarray}\label{0.1}
i\hbar\p_t\Psi=\hat H \Psi, \qquad \hat H= c\alpha^i\hat p_i+mc^2\beta.
\end{eqnarray}
Then $\hat H$ may be interpreted as the Hamiltonian. If we pass
from the Schr\"odinger to Heisenberg picture, time derivative of
an operator is $i\hbar\dot a=[a, H]$. For the basic operators of
the Dirac theory we obtain
\begin{eqnarray}\label{0.2}
\dot x_i=c\alpha_i, \qquad i\hbar\dot\alpha_i=2(cp_i-H\alpha_i),
\qquad \dot p_i=0.
\end{eqnarray}
Some immediate consequences of these equations are enumerated below.

\begin{itemize}

\item The wrong balance of the number of degrees of freedom.
According to the first equation from (\ref{0.2}), the operator
$c\alpha^i$ represents velocity of the "center of charge" $x^i$
\cite{sch}. Then the physical meaning of the operator $p^i$ became
rather obscure in both the semiclassical and the RQM framework.

Various approaches to the problem has been considered in the literature. Schr\"odinger noticed \cite{sch} that besides
the center of charge, $x$, in the Dirac theory we can construct the "center-of-mass" operator $\tilde x_i=x_i+\frac12
i\hbar cH^{-1}\alpha_i$ in such a way, that $p^i$ turns out to be the mechanical momentum for $\tilde x$, $\dot{\tilde
x}\sim p$ (various versions of this operator have been discussed in the works \cite{pr, 7, nw}). Following this way,
Schr\"odinger assumed the naive interpretation of the Dirac electron as a kind of composed system (we return to this
subject in Subsection 6.1). In contrast, Foldy and Wouthuysen \cite{7} assumed that $x^i$ does not correspond to an
observable quantity.

\item Free electron follows complicated trembling trajectory. The
equations (\ref{0.2}) can be solved, with the result for $x^i(t)$
being \cite{sch, dirbook}
\begin{eqnarray}\label{2.13.9}
x^i=a^i+bp^it+c^i\mbox{exp}(-\frac{2iH}{\hbar}t).
\end{eqnarray}
The trajectory is a superposition of the rectilinear motion along a straight line, $a_i+bp_it$, and the rapid
oscillations with higher frequency $\frac{2H}{\hbar}\sim\frac{2mc^2}{\hbar}$. The oscillator motion is called {\it
Zitterbewegung}. Schr\"odinger had compared spin with the angular momentum associated to the {\it Zitterbewegung}. He
found \cite{sch} that they differ by the factor $2$ and concluded that spin can not be identified with the {\it
Zitterbewegung}. In our model, we would be able to construct the variables for which the identification turns out to be
possible, see Subsection 6.1.

\item Since the velocity operator $c\alpha^i$ has eigenvalues $\pm
c$, we conclude that a measurement of a component of the velocity
of a free electron is certain to lead to the result $\pm c$.
Besides, since the operators $\alpha^i$ do not commute, components
of velocity in different directions can not be instantaneously
measured.

\end{itemize}

In view of this, one may assume that the Dirac equation does not admit the RQM interpretation. Other possibility might
be that the basic operator $\hat x$ appeared in the Dirac equation do not correspond to the physically observable
quantity. This possibility is supported by the seminal work of Foldy and Wouthuysen \cite{7}, where they have
constructed (in a Lorentz non-covariant manner) the position operator with reasonable properties. In particular, it
obeys the equation $\dot {\bf X}^i=\frac{p^i}{p^0}$. This leads to further complications, as the Dirac equation gives
no evidence which of these two operators should be identified with the position of an electron. We return to the
subject in Section 5, where we propose the Lorentz-covariant classical-mechanical analog for the Foldy-Wouthuysen
operator.

To understand the controversial properties of the one-particle
Dirac equation, it would be desirable to have at our disposal the
spinning-particle model which leads to the Dirac equation in the
course of canonical quantization. We construct and discuss the
analogous model of such a kind in this work.  It shows the same
properties as those of the Dirac equation in the RQM
interpretation. Analyzing the present model, we have been able to
identify the origin of the problems, see Section 7. The modified
model which turns out to be free of the problems mentioned above
has been proposed in the recent work \cite{aad1}.

The work is organized as follows. The operators of the Dirac theory which are associated with spin are the
$\Gamma^\mu$\,-matrices as well as the Lorentz generators $\Gamma^{\mu\nu}$. Their commutators can be identified with
the Poisson-brackets of angular momentum of five-dimensional space with the metric $\eta=(-,+,+,+,-)$ \cite{aad1}. So
the space can be taken as the underlying configuration space for the description of spin. As the number of independent
angular-momentum components is less than dimension of the spin phase space, dynamics of spin should be restricted to an
appropriate subspace which we call the spin surface\footnote{For the case of non-relativistic spin \cite{aad2}, the
surface can be identified with the group manifold $SO(3)$. This has the natural structure of a fiber bundle with base
$S^2$ and fiber $SO(2)$ The spin is identified with coordinates of the base. We are grateful to Prof. A. Nersessian for
the comments on this subject.} \cite{aad1}. The corresponding configuration space turns out to be the anti-de Sitter
space. To make the work self-consistent, we present the construction of the spin surface in Section 2.

Hamiltonian formulation of the model with spin sector of such a
kind has been announced in \cite{aad3}. In the Sections 3, 4 and 5
we carry out the detailed analysis of Hamiltonian and Lagrangian
formulations of the model and analyze it on the classical as well
as on the quantum level. In the Section 6 the classical equations
of motion are integrated and discussed in details. Section 7 is
left for the conclusions.

\section{Algebraic construction of relativistic spin surface}
We start from the Dirac equation written in the
manifestly-covariant form
\begin{eqnarray}\label{1.1}
(\hat p_\mu\Gamma^\mu+mc)\Psi(x^\mu)=0,
\end{eqnarray}
where $\hat p_\mu=-i\hbar\partial_\mu$. We use the representation
with hermitian $\Gamma^0$ and antihermitian $\Gamma^i$
\begin{eqnarray}\label{1.1 2}
\Gamma^0= \left(
\begin{array}{cc}
1& 0\\
0& -1
\end{array}
\right), \quad \Gamma^i= \left(
\begin{array}{cc}
0& \sigma^i\\
-\sigma^i& 0
\end{array}
\right),
\end{eqnarray}
then $[\Gamma^\mu, \Gamma^\nu]_{+}=-2\eta^{\mu\nu}$, $\eta^{\mu\nu}=(- + + +)$, and $\Gamma^0\Gamma^i$, $\Gamma^0$ are
the Dirac matrices $\alpha^i$, $\beta$ \cite{dirbook}. We take the classical counterparts of the operators $\hat x^\mu$
and $\hat p_\mu=-i\hbar\partial_\mu$ in the standard way, which are $x^\mu$, $p^\nu$, with the Poisson brackets
$\{x^\mu, p^\nu\}=\eta^{\mu\nu}$.

Let us look for the classical variables that could produce the
$\Gamma$\,-matrices. According to the canonical quantization
paradigm, the classical variables, say $z^\alpha$, corresponding
to the Hermitian operators $\hat z^\alpha$ should obey the
quantization rule
\begin{eqnarray}\label{0}
[\hat z^\alpha , \hat z^\beta]=i\hbar\left.\{z^\alpha , z^\beta
\}\right|_{z\rightarrow\hat z}.
\end{eqnarray}
In this equation, $[ ~ , ~]$ is the commutator of the operators and $\{ ~ , ~ \}$ stands for the classical
bracket\footnote{It is the Poisson (Dirac) bracket in a theory without (with) second-class constraints.}. To avoid the
operator-ordering problems, we will consider only the sets of operators which form the Lie algebra, $[\hat z^\alpha ,
\hat z^\beta]=c^{\alpha\beta}{}_\gamma\hat z^\gamma$. So our first task is to study the algebra of $\Gamma$\,-matrices.
We note that commutators of $\Gamma^\mu$ do not form closed Lie algebra, but produce $SO(1, 3)$ Lorentz generators
\begin{eqnarray}\label{1.2}
[\Gamma^\mu, \Gamma^\nu]=-2i\Gamma^{\mu\nu},
\end{eqnarray}
where
$\Gamma^{\mu\nu}\equiv\frac{i}{2}(\Gamma^\mu\Gamma^\nu-\Gamma^\nu\Gamma^\mu)$.
The set $\Gamma^\mu$, $\Gamma^{\mu\nu}$ forms a closed algebra.
Besides the commutator (\ref{1.2}), we have
\begin{eqnarray}\label{1.3}
[\Gamma^{\mu\nu}, \Gamma^\alpha]=2i(\eta^{\mu\alpha}\Gamma^\nu-\eta^{\nu\alpha}\Gamma^\mu), \qquad \qquad \cr
[\Gamma^{\mu\nu}, \Gamma^{\alpha\beta}]=2i(\eta^{\mu\alpha}\Gamma^{\nu\beta}- \eta^{\mu\beta}\Gamma^{\nu\alpha}-
\eta^{\nu\alpha}\Gamma^{\mu\beta}+\eta^{\nu\beta}\Gamma^{\mu\alpha}).
\end{eqnarray}
The algebra can be identified with $SO(2, 3)$ Lorentz algebra with
generators $\hat J^{AB}$:
\begin{eqnarray}\label{1.4}
[\hat J^{AB}, \hat J^{CD}]=2i(\eta^{AC}\hat J^{BD}-\eta^{AD}\hat
J^{BC}- \cr \eta^{BC}\hat J^{AD}+\eta^{BD}\hat J^{AC}),
\end{eqnarray}
assuming $\Gamma^\mu\equiv \hat J^{5\mu}$, $\Gamma^{\mu\nu}\equiv
\hat J^{\mu\nu}$.

To reach the algebra starting from a classical-mechanics model, we
introduce ten-dimensional "phase" space of the spin degrees of
freedom, $\omega^A$, $\pi^B$, equipped with the Poisson bracket
$\{\omega^A, \pi^B\}=\eta^{AB}$.
Consider the inner angular momentum
\begin{eqnarray}\label{1.6}
J^{AB}\equiv 2(\omega^A\pi^B-\omega^B\pi^A).
\end{eqnarray}
Poisson brackets of these quantities form the algebra
\begin{eqnarray}\label{1.4.1}
\{J^{AB}, J^{CD}\}_{PB}=2(\eta^{AC} J^{BD}-\eta^{AD}J^{BC}- \cr
\eta^{BC}J^{AD}+\eta^{BD}J^{AC}).
\end{eqnarray}
Comparing (\ref{1.4.1}) with (\ref{1.4}) we conclude that the
operators $\Gamma^\mu$, $\Gamma^{\mu\nu}$ could be obtained by
quantization of $J^{AB}$.

Since $J^{AB}$ are the variables which we are interested in, we
try to take them as coordinates of the space $\omega^A, \pi^B$.
The Jacobian of the transformation $(\omega^A, \pi^B)\rightarrow
J^{AB}$ has rank equal seven\footnote{The rank has been computed
using the program: Wolfram Mathematica 8.}. So, only seven among
ten functions $J^{AB}(\omega, \pi)$, $A<B$, are independent
quantities. They can be separated as follows. By construction, the
quantities (\ref{1.6}) obey the identity
$\epsilon^{\mu\nu\alpha\beta}J^5{}_\nu J_{\alpha\beta}=0$, this
can be solved as
\begin{eqnarray}\label{1.61}
J^{ij}=(J^{50})^{-1}(J^{5i}J^{0j}-J^{5j}J^{0i}).
\end{eqnarray}
Hence we can take $J^{5\mu}$, $J^{0i}$ as the independent variables. We could complete the set up to a base of the
phase space $(\omega^A, \pi^B)$ adding three more coordinates, for instance $\omega^3$, $\omega^5$, $\pi^5$. Quantizing
the complete set we obtain, besides the desired operators $\hat J^{5\mu}, \hat J^{0i}$, some extra operators
$\hat\omega^3$, $\hat\omega^5$, $\hat\pi^5$. They are not present in the Dirac theory, and are not necessary for
description of spin. So we need to reduce the dimension of our space from ten to seven imposing three constraints which
we denote $T_a(\omega, \pi)=0$, $a=3, 4, 5$. We require the surface defined by the constraints be invariant under
action of the angular-momentum generators, that is
\begin{eqnarray}\label{1.63.1}
\{T_a(\omega, \pi), J^{AB}\}=0.
\end{eqnarray}

The only quadratic $SO(2, 3)$\,-invariants which can be
constructed from $\omega^A$, $\pi^B$ are $\omega^A\omega_A$,
$\omega^A\pi_A$ and $\pi^A\pi_A$. So we restrict our model to live
on the surface defined by the equations
\begin{eqnarray}\label{1.7}
T_3\equiv\pi^A\pi_A+a_3=0;
\end{eqnarray}
\begin{eqnarray}\label{1.71}
T_4\equiv\omega^A\omega_A+a_4=0, \quad T_5\equiv\omega^A\pi_A=0,
\end{eqnarray}
where $a_3$, $a_4$ are positive numbers\footnote{The positivity guarantees the causal dynamics of our particle, see Eq.
(\ref{ca}).}. It is called the spin surface.

The first equation from (\ref{1.71}) states that the configuration
space of the spin degrees of freedom is anti-de Sitter space.

In the Hamiltonian formulation, the equations $T_a=0$ appear as
the Dirac constraints. So, we classify them in accordance with
their algebraic properties with respect to the Poisson bracket.
The brackets read
\begin{eqnarray}\label{1.66}
\{T_3, T_4\}=-4T_5, \qquad \{T_3, T_5\}=-2T_3+2a_3, \cr \{T_4, T_5\}=2T_4-2a_4. \qquad \qquad \qquad
\end{eqnarray}
Taking the combination
\begin{eqnarray}\label{1.66.1}
\tilde T_3\equiv T_3+\frac{a_3}{a_4}T_4,
\end{eqnarray}
we have the algebra
\begin{eqnarray}\label{1.66.2}
\{\tilde T_3, T_4\}=-4T_5, \qquad \{\tilde T_3, T_5\}=-2T_3+2\frac{a_3}{a_4}T_4, \cr \{T_4, T_5\}=-2a_4+2T_4, \qquad
\qquad \qquad
\end{eqnarray}
that is Poisson bracket of $\tilde T_3$ with any constraint vanishes on the surface, while the Poisson brackets of
constraints $T_4$, $T_5$ form an invertible matrix on the surface. In the Dirac terminology, the set $\tilde T_3$ is
the first-class constraint, while $T_4$, $T_5$ is a pair of second-class constraints.

There are various reasons to take the functions $T_a$ invariant.

\par \noindent A) Consistency of canonical quantization of a system with
second-class constraints implies replacement the Poisson by the
Dirac bracket, the latter is constructed with help of the
constraints. For the momentum generators it reads ($i=4, 5$)
\begin{eqnarray}\label{1.32}
\{J^{AB}, J^{CD}\}_{DB}=\{J^{AB}, J^{CD}\}-\{J^{AB}, T_i\}\{T_i, T_j\}^{-1}\{T_j, J^{CD}\}.
\end{eqnarray}
If the surface is invariant, the second term on the r. h. s. vanishes, and the Dirac bracket of $J^{AB}$ coincides with
the Poisson bracket, Eq. (\ref{1.4.1}). So, as before, we have the desired algebra.

\par \noindent B) Presence the first-class constraint $\tilde T_3$ implies that we
deal with a theory with local symmetry. Generators of the symmetry are proportional to the constraints \cite{gt, aad4,
aadbook}. Suppose that the first-class constraint is not invariant, that is we have $\{\tilde T_3, J^{AB}\}_{PB}\ne 0$.
It should imply that the variables $J^{AB}$ are not inert under the local symmetry, $\delta J^{AB}\sim\{\tilde T_3,
J^{AB}\}_{PB}\ne 0$. Hence the non-invariant constraints would be responsible for gauging out some of the variables
$J^{AB}$, which is not under our interest now.

Let us discuss convenient parametrization of the spin surface. The matrix $\frac{\partial(J^{5\mu}, J^{0i}, T_4, T_5,
\omega^5)}{\partial(\omega^A, \pi^B)}$ has rank equal ten. So the quantities
\begin{eqnarray}\label{1.71.1}
J^{5\mu}, ~ J^{0i}, ~ T_4, ~ T_5, ~ \omega^5,
\end{eqnarray}
can be taken as coordinates of the space $(\omega^A$, $\pi^B)$.
The equation $J^{AB}=2(\omega^A\pi^B-\omega^B\pi^A)$ implies the
identity
\begin{eqnarray}\label{1.68}
J^{AB}J_{AB}=8[(\omega^A)^2(\pi^B)^2-(\omega^A\pi_A)^2]= \cr
8[(T_4-a_4)(T_3-a_3)-(T_5)^2], \quad
\end{eqnarray}
then the constraint $T_3$ can be written in the coordinates
(\ref{1.71.1}) as follows:
\begin{eqnarray}\label{1.71.2}
T_3=\frac{(J^{AB})^2+8(T_5)^2}{8(T_4-a_4)}+a_3,
\end{eqnarray}
where $J^{ij}$ are given by Eq. (\ref{1.61}). Note that $T_3$ does not depend on $\omega^5$. On the hyperplane
$T_4=T_5=0$ it reduces to
\begin{eqnarray}\label{1.71.3}
-8a_4T_3=(J^{AB})^2-8a_3a_4=0.
\end{eqnarray}
Eq. (\ref{1.71.3}) states that the value of $SO(2, 3)$\,-Casimir
operator $(J^{AB})^2$ is equal to $8a_3a_4$. In quantum theory,
for the operators (\ref{1.4}), (\ref{1.3}) we have: $\hat
J^{AB}\hat J_{AB}=20\hbar^2$. So we can fix the product of our
parameters as
\begin{eqnarray}\label{main}
a_3a_4=\frac{5\hbar^2}{2}.
\end{eqnarray}

As we have pointed out above, the function $T_3$ represents generator of local symmetry. The coordinate $\omega^5$ is
not inert under the symmetry, $\delta\omega^5\sim\{T_3, \omega^5\}\ne 0$. Hence $\omega^5$ is gauge non-invariant
(hence non-observable) variable.

Summing up, we have restricted dynamics of spin on the surface (\ref{1.7}), (\ref{1.71}). If (\ref{1.71.1}) are taken
as coordinates of the phase space, the surface is the hyperplane $T_4=T_5=0$ with the coordinates $J^{5\mu}, J^{0i},
\omega^5$ subject to the condition (\ref{1.71.3}). Since $\omega^5$ is gauge non-invariant coordinate, we can discard
it. It implies that we can quantize $J^{5\mu}, J^{0i}$ instead of the initial variables $\omega^A$, $\pi^B$.

Following the canonical quantization paradigm, the variables must
be replaced by Hermitian operators\footnote{The matrices
$\Gamma^\mu$, $\Gamma^{\mu\nu}$ are Hermitian operators with
respect to the scalar product $(\Psi_1,
\Psi_2)=\Psi_1^\dagger\Gamma^0\Psi_2$.} with commutators
resembling the Poisson bracket
\begin{eqnarray}\label{1.71.5}
[ ~ , ~ ]=i\hbar\left.\{ ~  ,  ~ \}\right|_{J\rightarrow\hat J}.
\end{eqnarray}
Similarly to the case of $\Gamma$\,-matrices, brackets of the
variables $J^{5\mu}$, $J^{0i}$ do not form closed Lie algebra. The
non closed brackets are
\begin{eqnarray}\label{1.71.6}
\{J^{5i}, J^{5j}\}=\{J^{0i}, J^{0j}\}=-2J^{ij},
\end{eqnarray}
where $J^{ij}$ is given by Eq. (\ref{1.61}). Adding them to the
initial variables, we obtain the set $J^{AB}=(J^{5\mu}, J^{0i},
J^{ij})$ which obeys the desired algebra (\ref{1.4.1}).

According to Eqs. (\ref{1.4}), (\ref{1.4.1}) the quantization is achieved replacing the classical variables $J^{5\mu}$,
$J^{\mu\nu}$ on $\Gamma$\,-matrices\footnote{Replacing (\ref{1.61}) by an operator $\hat J^{ij}(\Gamma^\mu,
\Gamma^{0i})$ we arrange the operators $\Gamma$ in such a way, that $\hat J^{ij}(\Gamma)=\Gamma^{ij}$.}. We assume that
$\omega^A$ has a dimension of length, then $J^{AB}$ has the dimension of the Planck's constant. Hence the quantization
rule is
\begin{eqnarray}\label{1.9}
J^{5\mu}\rightarrow\hbar\Gamma^\mu, \quad
J^{\mu\nu}\rightarrow\hbar\Gamma^{\mu\nu}.
\end{eqnarray}

This implies that the Dirac equation (\ref{1.1}) can be produced
by the constraint
\begin{eqnarray}\label{1.8}
T_2\equiv p_\mu J^{5\mu}+mc\hbar=0.
\end{eqnarray}

\par
\noindent
\section{Hamiltonian formulation}
Our next task is to formulate the variational problem for our
spinning particle. As it has been discussed above, we need a
theory which implies the constraints (\ref{1.7}), (\ref{1.71}) and
(\ref{1.8}). Since they are written on the phase space, it is
natural to start from construction of an action functional in the
Hamiltonian formalism.

Hamiltonian action of non-singular (that is non-constrained) system reads $\int d\tau[P\dot Q-H(Q, P)]$. When the
theory is singular, Hamiltonian action acquires more complicated form \cite{dir, gt, aadbook}. To remind its structure,
as well as to justify our choice of the action (\ref{3.0}), (\ref{3.01}), we outline the Hamiltonian formulation for
singular Lagrangian theory of a special form. For our purposes it will be sufficient to discuss the Lagrangian
\begin{eqnarray}\label{4.1}
S=\int d\tau L(Q, \dot Q, e_a),
\end{eqnarray}
that is the configuration variables are divided on two groups, $Q$
and $e_a$, where $e_a$ enter into the action without derivatives.
We also suppose that
\begin{eqnarray}\label{4.2}
\det\frac{\partial ^2L}{\partial\dot Q\partial\dot Q}\ne 0.
\end{eqnarray}
Following the standard prescription, we construct Hamiltonian
formulation for the action (\ref{4.1}). Canonical momenta are
defined by
\begin{eqnarray}\label{4.3.1}
P=\frac{\partial L}{\partial\dot Q},
\end{eqnarray}
\begin{eqnarray}\label{4.4}
\pi_{ea}=\frac{\partial L}{\partial\dot e_a}=0.
\end{eqnarray}
The phase space $(Q, P; e_a, \pi_{ea})$ is equipped with canonical
(nondegenerated) Poisson bracket. Nonvanishing brackets are $\{Q,
P\}_{PB}=1$, $\{e_a, \pi_{ea}\}_{PB}=1$.

According to Eq. (\ref{4.4}), in the theory there are the primary constraints $\pi_{ea}=0$. Due to the condition
(\ref{4.2}), Eqs. (\ref{4.3.1}) can be resolved with respect to $\dot Q$, $\dot Q=f(Q, P, e_a)$. Using these
expressions, we construct complete Hamiltonian according to the standard rule
\begin{eqnarray}\label{4.5}
H(Q, P, e_a, \pi_{ea}, \lambda_{ea})= \qquad \qquad \qquad \qquad
\qquad \cr \left.\left[P\dot Q+\pi_{ea}\dot e_a-L(Q, \dot Q,
e_a)\right]\right|_{\dot Q=f(Q, P, e),
\pi_{ea}=0}+\lambda_{ea}\pi_{ea},
\end{eqnarray}
where $\lambda_{ea}(\tau)$ are the Lagrangian multipliers for the
primary constraints (\ref{4.4}).  Given complete Hamiltonian and
the Poisson brackets, the temporal evolution of any quantity $A$
is given by the equation $\dot A=\{A, H\}_{PB}$. In particular,
the basic variables obey the Hamiltonian equations
\begin{eqnarray}\label{4.6}
\dot Q=\{Q, H\}_{PB}, \quad \dot P=\{P, H\}_{PB}, \quad  \dot e_a=\lambda_{ea}, \quad  \dot\pi_{ea}=0.
\end{eqnarray}
The condition of preservation in time for the primary constraints
generally implies the secondary constraints denoted by $T_a$
\begin{eqnarray}\label{4.7}
\dot\pi_{ea}=\{\pi_{ea}, H\}_{PB}=-\frac{\partial H}{\partial
e_a}\equiv T_a=0.
\end{eqnarray}
Dynamics of the system can be equivalently obtained starting from
the Hamiltonian action functional
\begin{eqnarray}\label{4.8}
S_H=\int d\tau\left[P\dot Q+\pi_{ea}\dot e_a-H(Q, P,e_a, \pi_{ea},
\lambda_{ea})\right].
\end{eqnarray}
Performing variation of the action with respect to all the
variables $(Q, P, e_a, \pi_{ea}, \lambda_{ea})$, we obtain the
complete set of equations governing the dynamics, (\ref{4.4}),
(\ref{4.6}), (\ref{4.7}).

The procedure can be inverted. Given Hamiltonian action (\ref{4.8}), we can restore the corresponding Lagrangian
formulation. In the process, we can omit $\pi_{ea}$ because $e_a$ enter into the formulation without derivatives.
Assuming that the Hamiltonian (\ref{4.5}) is at most quadratic function of the momenta $P$, we write it in the form
$H=\frac12 PG(Q, e)P+g(Q, e)$. We solve the Hamiltonian equations $\dot Q=GP$ with respect to $P$. The solution reads
$P=\tilde G\dot Q$, where $\tilde G$ is the inverse matrix for $G$. We substitute these $P$ back into Eq. (\ref{4.8}),
obtaining the Lagrangian action
\begin{eqnarray}\label{4.15}
S=\int d\tau\left[\frac12\dot Q\tilde G\dot Q-g(Q, e)\right].
\end{eqnarray}
Applying the hamiltonization procedure to the action, we expect to
arrive back at the Hamiltonian formulation (\ref{4.8}). We do this
for our model in Sect. 5.

We are interested in to construct the variational problem which implies the constraints (\ref{1.7}), (\ref{1.71}) and
(\ref{1.8}). The analysis made above suggests to take the following Hamiltonian action\footnote{The Hamiltonian
(\ref{3.01}) is slightly different from those of the work \cite{aad3}. It will lead to more simple Lagrangian. Both
theories have the same physical sector.}
\begin{eqnarray}\label{3.0}
S_H=\int d \tau \left( p_{\mu}\dot x^{\mu} + \pi_A \dot \om^A +
\pi_{e_l} \dot e_l - H \right),
\end{eqnarray}
with the Hamiltonian constructed on the base\footnote{Peculiar
property of the model is that the equation $T_5=0$ appears as the
third-stage constraint, from the condition of preservation in time
of the constraint $T_4$. So, it is not necessary to include $T_5$
into the Hamiltonian.} of constraints $T_2$, $T_3$ and $T_4$
\begin{eqnarray}\label{3.01}
H=\frac{e_l}{2}T_l+\lambda_{e_l}\pi_{e_l}= \qquad \qquad \qquad
\qquad \cr \frac{e_2}{2}(p_{\mu}J^{5\mu}+mc\hbar)+
\frac{e_3}{2}[(\pi^A)^2+a_3]+\frac{e_4}{2}[(\om^A)^2+a_4]+\lambda_{e_l}\pi_{e_l}.
\end{eqnarray}
Variation of $S_H$ with respect to momenta gives us the Hamiltonian equations for position variables,
\begin{eqnarray}\label{3.001}
\frac{\de S_H}{\de p_{\mu}}=0 & \Rightarrow & \dot
x^{\mu}=\frac{e_2}{2} J^{5 \mu} = e_2(\om^5\pi^{\mu}
-\pi^5\om^{\mu} ),
\end{eqnarray}
\begin{eqnarray}\label{3.0012}
\frac{\de S_H}{\de \pi_{\mu}}=0 & \Rightarrow &
\dot \omega^{\mu}= e_3\pi^{\mu} + e_2\omega^5 p^{\mu},
\end{eqnarray}
\begin{eqnarray}\label{3.0013}
\frac{\de S_H}{\de \pi_5}=0 &\Rightarrow & \dot
\omega^5 = e_3\pi^5 + e_2p_{\mu}\om^{\mu},
\end{eqnarray}
\begin{eqnarray}\label{3.0014}
\frac{\de S_H}{\de \pi_{e_l}}=0 & \Rightarrow & \dot e_l=\lambda_{e_l}.
\end{eqnarray}
Variation of $S_H$ with respect to the position variables gives the equations for momenta,
\begin{eqnarray}\label{3.002}
\frac{\de S_H}{\de x^{\mu}}=0 & \Rightarrow & \dot p_{\mu}=0,
\end{eqnarray}
\begin{eqnarray}\label{3.0021}
\frac{\de S_H}{\de \om^{\mu}}=0 & \Rightarrow & \dot \pi^{\mu}=
e_2 \pi^5 p^{\mu} - e_4 \om^{\mu},
\end{eqnarray}
\begin{eqnarray}\label{3.0022}
\frac{\de S_H}{\de \om^5}=0 & \Rightarrow & \dot \pi^5 = e_2
p_{\mu} \pi^{\mu} - e_4 \omega^5.
\end{eqnarray}
Besides we obtain the primary constraints $\pi_{e_l}=0$, which
appear from variation with respect to the Lagrange multipliers
$\lambda_{e_l}$. At last, the variation of $S_H$ with respect to
$e_l$ gives a part of the desired constraints of the theory,
\begin{eqnarray}\label{3.003}
\frac{\de S_H}{\de e_2}=0 & \Rightarrow & p_{\mu}J^{5 \mu}+mc
\hbar=0,
\end{eqnarray}
\begin{eqnarray}\label{3.0031}
\frac{\de S_H}{\de e_3}=0 & \Rightarrow & (\pi^A)^2 + a_3=0,
\end{eqnarray}
\begin{eqnarray}\label{3.0032}
\frac{\de S_H}{\de e_4}=0 & \Rightarrow & (\om^A)^2+a_4 = 0.
\end{eqnarray}
Preservation in time of the constraint $(\om^A)^2+a_4=0$ gives the
constraint $T_5=\om^A\pi^A=0$, which preservation, in turn, leads
to the $e_4-\frac{a_3}{a_4}e_3=0$. To see this, we use equations
of motion (\ref{3.0012}), (\ref{3.0013}), (\ref{3.0021}) and
(\ref{3.0022}),
\begin{eqnarray}\label{3.00310}
[(\om^A)^2 +a_4]\dot{} = 2\dot \om^A \om^A = 2e_3\om^A \pi^A = 0 \Rightarrow \om^A \pi^A = 0; \\ \label{3.00320} (\om^A
\pi^A) \dot{} = e_3(\pi^A)^2- e_4(\om^A)^2 = 0 \Rightarrow e_4-\frac{a_3}{a_4}e_3=0.
\end{eqnarray}
Time derivative of the constraint $e_4-\frac{a_3}{a_4}e_3=0$ determines the Lagrangian multiplier $\lambda_{e4}$,
$\lambda_{e4}=\frac{a_3}{a_4}\lambda_{e3}$. Preservation in time of the constraints $T_2=0$ and $T_3=0$ gives no new
equations.

In the result, we have two pairs of second-class constraints, $e_4-\frac{a_3}{a_4}e_3=0$, $\pi_{e_4}=0$ and $T_4=0$,
$T_5=0$, while $\pi_{e_2}=0$, $\pi_{e_3}=0$, $T_2=0$ and $\ti T_3 = T_3 + \frac{a_3}{a_4}T_4=0$ represent first-class
constraints. The first two of them are primary, and according to the general theory \cite{dir, gt, aadbook, aad4} it
indicates on invariance of $S_H$ with respect to two-parameter group of local transformations.

One of them is the well-known reparametrization transformation,
its infinitesimal form is
\begin{eqnarray}\label{ri1}
\de Y=\alpha\dot Y, \quad \de e_l=(\alpha e_l)\dot{}, \quad \de \pi_{e_l}=0, \quad \de\lambda_{e_l}=(\de e_{l})\dot{},
\end{eqnarray}
where $Y =(x^{\mu}, p_{\mu}, \om^A, \pi_B)$ and
$\alpha=\alpha(\tau)$ is an arbitrary function. Variation of $S_H$
with respect to (\ref{ri1}) is equal to the total derivative
\begin{eqnarray}
\de S_H =\int d\tau[\alpha(p_{\mu}\dot x^{\mu}+\pi_A \om^A-\frac{1}{2}e_lT_l)]^..
\end{eqnarray}
The other symmetry is given by\footnote{It is Hamiltonian
counterpart of the Lagrangian symmetry which will be discussed
below, see Eqs. (\ref{3.2})-(\ref{3.3.1}).}
\begin{eqnarray}\label{s1}
\delta  x^{\mu}=0, \quad \delta p_{\mu}=0,
\end{eqnarray}
\begin{eqnarray}\label{s2}
\delta\omega^A =\epsilon e_3\pi^A, \quad \delta\pi^A=-\epsilon e_3\omega^A,
\end{eqnarray}
\begin{eqnarray}\label{s3}
\delta e_2=0, \quad  \delta e_3=(\epsilon e_3)\dot{}, \quad  \delta e_4=(\epsilon e_4)\dot{},
\end{eqnarray}
\begin{eqnarray}\label{s4}
\de \lambda_{e_l} = (\de e_{l})\dot {}, \quad
 \de \pi_{e_l} = 0,
\end{eqnarray}
Variation of $S_H$ is equal to a total derivative as well. \par

\noindent{\it Comment.} In the Berezin-Marinov model \cite{bm75} the Dirac equation is implied by the supersymmetric
gauge transformations. In our model the Dirac equation is associated with the $\varepsilon$\,-symmetry. So, it
represents the bosonic analogue of BM supersymmetric transformations.

According to general theory \cite{dir, gt, aadbook}, the local symmetries indicate on two-parametric degeneracy in
solutions to equations of motion. Indeed, we note that $\lambda_{e_2}(\tau)$ and $\lambda_{e_3}(\tau)$ cannot be
determined neither with the system of constraints nor with the dynamical equations. As a consequence (see Eq.
(\ref{3.0014})), the variables $e_2$ and $e_3$ turn out to be the arbitrary functions as well. Since they enter into
the equations for $x^{\mu}, \om^A$, and $\pi_A$, general solution for these variables contains, besides the arbitrary
integration constants, the arbitrary functions $e_a$. The variables with ambiguous evolution have no physical meaning
\cite{gt}. Hence our next task is to find candidates for observables, which are variables with unambiguous dynamics.
Equivalently, we can look for the gauge-invariant variables.

The only unambiguous among the initial variables is $p^\mu$, see
Eq. (\ref{3.002}). $x^\mu$ has one-parameter ambiguity due to
$e_2$, while $\omega^A$ and $\pi^B$ have two-parameter ambiguity
due to $e_2$ and $e_3$. Inspection of equations of motion for
$\omega^A$ and $\pi^B$ allows us to construct more quantities with
one-parameter ambiguity, they turn out to be the angular-momentum
components
\begin{eqnarray}\label{3b.6}
\dot J^{\mu \nu} = e_2(p^{\mu}J^{5 \nu}- p^{\nu}J^{5 \mu}), \\
\label{3b.61} \dot J^{5 \mu} = e_2p_{\nu}J^{\nu \mu}. \qquad \quad
\end{eqnarray}
We also point out that $J^{AB}$ is invariant under the $\varepsilon$\,-transformation (\ref{s2}). The constraints
(\ref{1.7}) and (\ref{1.71}) determine the square of the angular-momentum tensor, see Eq. (\ref{1.71.3}). Besides, they
guarantee that $J^{5\mu}$ is the time-like vector
\begin{eqnarray}\label{k4}
(J^{5 \mu})^2=-4(a_3(\om^5)^2+a_4(\pi^5)^2)<0,
\end{eqnarray}
for positive values of $a_3$, $a_4$.

To proceed further, it is instructive to compare equations for $x^\mu$, $J^{\mu\nu}$ and $J^{5\mu}$ with those for
$x^\mu$ of spinless relativistic particle (see, for example, \cite{bru2, bru3}). If we use the parametric
representation $x^\mu(\tau)=(ct(\tau), x^i(\tau))$ for the trajectory $x^i(t)$, the spinless particle can be described
by the action
\begin{eqnarray}\label{k}
S=\int d\tau(\frac{1}{2 e}\dot x^2-\frac{e}{2}m^2c^2).
\end{eqnarray}
It implies the Hamiltonian equations
\begin{eqnarray}\label{k1}
\dot x^{\mu} = e p^{\mu}, \qquad \dot p_{\mu}= 0,
\end{eqnarray}
as well as the constraint
\begin{eqnarray}\label{k2}
p^2 + m^2c^2 = 0.
\end{eqnarray}
The auxiliary variable $e$ enters into general solution for
$x^\mu(\tau)$ as an arbitrary function. The ambiguity reflects the
freedom in the choice of parametrization for the particle
trajectory
\begin{equation}
\begin{array}{c}
\tau\rightarrow\tau'=\tau+\alpha, \qquad \qquad \\
x^\mu(\tau)\rightarrow x'{}^\mu(\tau')=x^\mu(\tau), \quad \\
e(\tau)\rightarrow e'(\tau')=(1+\dot\alpha)e(\tau).
\end{array}
\mbox{then} ~
\begin{array}{c}
\de x^{\mu}=\alpha\dot x^{\mu}, \\
\quad\de e=(\alpha e)\dot{}.
\end{array}
\end{equation}
The action (\ref{k}) turns out to be invariant under the
reparametrizations.

By construction, the expression for the physical trajectory
$x^i(t)$ is obtained resolving the equation $x^0=x^0(\tau)$ with
respect of $\tau$, $\tau=\tau(x^0)$, then $x^i(t)\equiv
x^i(\tau(x^0))$. The last equality implies
\begin{eqnarray}\label{v.1}
\frac{dx^i}{dt}= c\frac{\dot x^i}{\dot x^0} = c\frac{p^i}{p^0}.
\end{eqnarray}
As it should be, the physical coordinate $x^i(t)$ has unambiguous evolution.

To see physical meaning of the constraint (\ref{k2}), we take
square of Eq. (\ref{v.1}) and use (\ref{k2}) to estimate the
particle's speed
\begin{eqnarray}
\left(\frac{d x^i}{dt}\right)^2=c^2 \frac{(p^i)^2}{(p^i)^2+m^2
c^2}\Rightarrow\left(\frac{d x^i}{dt}\right)^2<c^2.
\end{eqnarray}
Hence the constraint (\ref{k2}) guarantees that the particle's
speed can not exceed the speed of light.

The same result can be reproduced in the Lagrangian formulation. Indeed, variation of the action (\ref{k}) with respect
to $e$ implies that $x^\mu$ is the time-like vector, $(\dot x^\mu)^2=-e^2m^2c^2<0$. This also allows us to estimate the
particle's speed
\begin{equation}
\left(\frac{dx^i}{dt}\right)^2=c^2\left(\frac{\dot x^i}{\dot
x^0}\right)^2=c^2\left(1-\frac{e^2m^2c^2}{(\dot x^0)^2}\right)
\Rightarrow \left(\frac{d x^i}{dt}\right)^2<c^2.
\end{equation}

Let us return to the spinning particle. We note both $x^\mu$ and $J^{AB}$ are invariant under
$\varepsilon$\,-transformations (\ref{s1}), (\ref{s2}). So the ambiguity presented in equations of motion
(\ref{3.001}), (\ref{3b.6}) and (\ref{3b.61}) is due to the reparametrization symmetry (\ref{ri1}). In accordance with
this observation, we can assume that the functions $x^\mu(\tau)$, $J^{AB}(\tau)$ represent the physical variables
$x^i(t)$ and $J^{AB}(t)$ in the parametric form. Then equations of motion for the physical variables read
\begin{equation}\label{3b.4}
\frac{dx^i}{dt}=c\frac{J^{5 i}}{J^{50}},
\end{equation}
\begin{eqnarray}\label{3b.7}
\frac{dJ^{\mu\nu}}{dt}=c\frac{\dot J^{\mu\nu}}{\dot x^0}=
2c\frac{p^\mu J^{5\nu}-p^\nu J^{5\mu}}{J^{50}}, \\
\label{3b.71}  \frac{dJ^{5\mu}}{dt}=c\frac{\dot J^{5\mu}}{\dot x^0}=2c\frac{p_\nu J^{\nu\mu}}{J^{50}}.
\end{eqnarray}
As it should be, they are unambiguous. General solution to these
equations will be obtained in Sect. 6.

Although there is no the mass-shell constraint $p^2+ m^2c^2=0$ in
our model, our particle's speed cannot exceed the speed of light.
To see this, we take square of Eq. (\ref{3b.4}) and use the fact
that $J^{5\mu}$ is the time-like vector, see Eq. (\ref{k4}), to
estimate the particle's speed
\begin{eqnarray}\label{ca}
\left(\frac{dx^i}{dt}\right )^2=c^2\frac{(J^{5i})^2}{(J^{50})^2}=
c^2\frac{(J^{5i})^2}{(J^{5i})^2 + 4(a_3(\om^5)^2+a_4(\pi^5)^2)}
\Rightarrow \cr \left(\frac{dx^i}{dt} \right)^2<c^2. \qquad \qquad
\qquad \qquad \qquad
\end{eqnarray}
Note that the spinning particle has causal dynamics for both positive and negative values of $p^2$.

\section{Lagrangian formulation}
In this section we reconstruct the configuration-space formulation
of the theory (\ref{3.0}). We obtain various equivalent forms of
the Lagrangian action and analyze the Lagrangian equations. While
this is much less systematic procedure as compare with the Dirac
method, at the end we arrive at the essentially the same results
as those of the previous section. In what follows, we suppress the
four-dimensional indexes, for example, we write
$x^\mu\omega_\mu\equiv(x\omega)$. We introduce also the following
condensed notation: $P_a=(p_{\mu}, \pi_{\nu},\pi_5)$,
$Q^a=(x^{\mu}, \om^{\nu}, \om^5)$ and $a_2\equiv mc\hbar$.  Then
the Hamiltonian (\ref{3.01}) reads
\begin{eqnarray}\label{3.0041}
H=\frac{1}{2}P_{a}G^{ab}P_{b}+\frac12e_4(\omega^A)^2+\frac{a_l}{2}
e_l, \qquad l=2, 3, 4,
\end{eqnarray}
where $G^{ab}(\om^A, e_l)$ is a $9 \times 9$ non-singular matrix,
$\det G = e^8_2e_3(\om^5)^6 (\om^A)^2$, which is schematically
written as
\begin{eqnarray}\label{3.004}
G^{ab} = \left( \begin{array}{cccccc} 0_{(4\times 4)} &
e_2\om^5\eta^{\mu \nu}_{(4\times 4)} & e_2\om^{\nu}
_{(4\times 1)} \\
e_2\om^5\eta^{\mu \nu}_{(4\times 4)} & e_3\eta^{\mu \nu}
_{(4\times 4)} & 0_{(4\times 1)}
\\ e_2\om^{\mu}_{(1\times 4)} & 0_{(1\times 4)} & -{e_3}_{(1\times 1)}
\end{array} \right).
\end{eqnarray}
The notation $0_{(4\times 4)}$ indicates that the first block of
the matrix $G^{ab}$ is composed of the null $4\times 4$-matrix,
and so on.

To find the Lagrangian, we write Hamiltonian equations for the
position variables $Q^a$, $\dot Q^{a}=G^{ab}P_b$, and resolve them
with respect to $P_a$, $P_a=\tilde G_{ab}\dot Q^b$, where $\tilde
G$ is the inverse matrix for $G$. We substitute these $P_a$ back
into the Hamiltonian action (\ref{3.0}), which gives the desired
Lagrangian
\begin{eqnarray}\label{3.005}
L=\frac{1}{2}\tilde G_{ab} \dot Q^a \dot Q^b -\frac{a_l}{2}
e_l-\frac12e_4(\omega^A)^2.
\end{eqnarray}
Hence the problem of restoring the Lagrangian formulation is
reduced to obtaining the inverse matrix for $G^{ab}$. It reads
\begin{eqnarray}\label{3.007}
\tilde G_{ab}=\left(
\begin{array}{cccccc}
\tilde G_{\mu\nu} & -\frac{e_2\om^5}{e_3}\tilde G_{\mu\nu} &
\frac{1}{e_2(\om^A)^2}\om_{\nu} \\
-\frac{e_2\om^5}{e_3}\tilde G_{\mu\nu} & \frac{1}{e_3(\om^A)^2}\om_{\mu}
\om_{\nu} & -\frac{\om^5}{e_3(\om^A)^2} \om_{\nu} \\
\frac{1}{e_2(\om^A)^2}\om_{\mu} & -\frac{\om^5}{e_3(\om^A)^2} \om_{\mu} & \frac{(\om^5)^2}{e_3(\om^A)^2}
\end{array} \right),
\end{eqnarray}
where
\begin{eqnarray}\label{a.1.1}
\tilde G_{\mu\nu}=-\frac{e_3}{(e_2\om^5)^2}\left[\eta_{\mu\nu}-\frac{\om_{\mu} \om_{\nu}}{(\om^A)^2}\right].
\end{eqnarray}
It is invertible, with the inverse matrix being
\begin{eqnarray}\label{a.1.2}
G^{\mu\nu}\equiv-\frac{(e_2\om^5)^2}{e_3}\left[\eta^{\mu\nu}-\frac{\omega^\mu\omega^\nu}{(\omega^5)^2}\right].
\end{eqnarray}
We substitute these expressions into Eq. (\ref{3.005}) and obtain
the manifest form of our Lagrangian
\begin{eqnarray}\label{3.1}
L=-\frac{e_3}{2(e_2\om^5)^2}\dot x^2+\frac{1}{e_2\om^5}(\dot x\dot\om)+\frac{1}{2e_3(\omega^A)^2}\left(\om_A\dot
\om^A-\frac{e_3}{e_2\om^5}(\dot x\om)\right)^2- \cr \frac{e_2}{2}a_2-\frac{e_3}{2}a_3-\frac{e_4}{2}[(\om^A)^2+a_4].
\qquad \qquad \qquad
\end{eqnarray}
It is manifestly Poincare-invariant. The variables $\omega^5$,
$e_l$, are scalars under the Poincare transformations. The
remaining variables transform according to the rule
\begin{eqnarray}\label{2.1.1}
x'^{\mu}=\Lambda^\mu{}_\nu x^\nu+a^\mu, \qquad
\omega'^{\mu}=\Lambda^\mu{}_\nu\omega^\nu.
\end{eqnarray}

To discuss the local symmetries, it is convenient to rewrite $L$ in an equivalent form by rearranging its terms as
follows:
\begin{eqnarray}\label{3.3a}
L=\frac12\tilde G_{\mu\nu}Dx^\mu Dx^\nu+\frac{1}{2e_3}(\dot\om^A)^2-\frac{e_4}{2}(\om^A)^2-\frac{a_l}{2}e_l, \quad l=2,
3, 4.
\end{eqnarray}
$\tilde G_{\mu\nu}$ is the upper-left block of the matrix $\tilde G_{ab}$, see Eq. (\ref{a.1.1}). Besides, in Eq.
(\ref{3.3a}) it has been denoted
\begin{eqnarray}\label{3.3ab}
Dx^{\mu}=\dot x^{\mu}-I^{5\mu},
\end{eqnarray}
\begin{eqnarray}\label{3.3ac}
I^{5\mu}=\frac{e_2}{e_3}(\om^5\dot\om^{\mu}-\dot\om^5\om^{\mu}).
\end{eqnarray}
The Lagrangian implies the following expressions for conjugate momenta
\begin{equation}\label{ls.1}
p_{\mu}=\frac{\p L}{\p \dot x^{\mu}}=\tilde G_{\mu\nu} Dx^\nu=\tilde G_{\mu\nu}(\dot x^\nu-I^{5\nu}),
\end{equation}
\begin{equation}\label{ls.2}
\pi_{\mu}=\frac{\p L}{\p\dot\om^{\mu}}=\frac{1}{e_3}(\dot\omega_\mu-e_2\omega^5p_\mu),
\end{equation}
\begin{equation}\label{ls.3}
\pi^5=\frac{\p L}{\p\dot\om_5}=\frac{1}{e_3}(\dot\omega^5-e_2(\omega p)).
\end{equation}

Local symmetries of the theory form the two-parameter group. The first is the reparametrization symmetry with the
parameter $\alpha(\tau)$
\begin{eqnarray}\label{rp1}
\delta_\alpha x^{\mu}=\alpha\dot x^{\mu}, \qquad \delta_\alpha\om^A=\alpha\dot\om^A, \qquad \delta_\alpha e_l=(\alpha
e_l)\dot{},
\end{eqnarray}
The variations (\ref{rp1}) imply
\begin{eqnarray}
\delta\tilde G_{\mu\nu}=(\alpha\tilde G_{\mu\nu})\dot{}-2\dot\alpha\tilde G_{\mu\nu}, \quad \de Dx^{\mu}=(\alpha
Dx^{\mu})\dot{}, \cr \de \left(\frac{1}{2e_3}(\dot\om^A)^2 \right) = \left(\alpha\frac{1}{2e_3}(\dot\om^A)^2
\right)\dot{}, \qquad \quad \cr \de(e_4[(\om^A)^2+a_4])=(\alpha e_4[(\om^A)^2+a_4])\dot{}.
\end{eqnarray}
Using these equalities it is easy to verify that $\de L =(\alpha L)\dot{}$.

From equations (\ref{s1})-(\ref{s3}), (\ref{ls.2}) and (\ref{ls.3}) we expect that the second symmetry would be
\begin{eqnarray}\label{3.2}
\delta_\epsilon x^{\mu}=0,
\end{eqnarray}
\begin{eqnarray}\label{3.3}
\delta_\epsilon\omega^\mu=\epsilon\dot\omega^\mu-\epsilon e_2\omega^5 p^\mu, \qquad
\delta_\epsilon\omega^5=\epsilon\dot\omega^5-\epsilon e_2(\omega p),
\end{eqnarray}
\begin{eqnarray}\label{3.3.1}
\delta_\epsilon e_2=0, \quad  \delta_\epsilon e_3=(\epsilon e_3)\dot{}, \quad \delta_\epsilon e_4=(\epsilon e_4)\dot{}.
\end{eqnarray}
To simplify the computations, we combine $\delta_\epsilon$  and $\delta_\alpha$,
$\delta=\delta_\epsilon+\delta_\alpha$, with $\alpha=-\epsilon$. The combination reads
\begin{eqnarray}\label{ls.4}
\delta x^{\mu}=-\epsilon\dot x^\mu,
\end{eqnarray}
\begin{eqnarray}\label{ls.5}
\delta\omega^\mu=-\epsilon e_2\omega^5 p^\mu, \qquad \delta\omega^5=-\epsilon e_2(\omega p),
\end{eqnarray}
\begin{eqnarray}\label{ls.6}
\delta e_2=-(\epsilon e_2)\dot{}, \quad  \delta e_3=\delta e_4=0.
\end{eqnarray}
where $p^\mu$ is given by (\ref{ls.1}). These transformations imply
\begin{eqnarray}\label{ls.7}
\delta(\omega^A)^2=0,
\end{eqnarray}
\begin{eqnarray}\label{ls.8}
\delta Dx^\mu=-(\epsilon Dx^\mu)\dot{}-\epsilon e_2\left[\omega^5\left(\frac{\dot\omega^\mu}{e_3}\right)^.-\right. \cr
\left.\left(\frac{\dot\omega^5}{e_3}\right)^.\omega^\mu\right]-\frac{e_2}{e_3}\delta(\omega^5\dot\omega^\mu-\dot\omega^5\omega^\mu),
\end{eqnarray}
\begin{eqnarray}\label{ls.9}
\delta\tilde G_{\mu\nu}=2\left[\frac{(\epsilon e_2)\dot{}}{e_2}+\frac{\epsilon e_2(\omega p)}{\omega^5}\right]\tilde
G_{\mu\nu}-\frac{\epsilon e_3}{e_2\omega^5(\omega^A)^2}(p_\mu\omega_\nu+p_\nu\omega_\mu),
\end{eqnarray}
then
\begin{eqnarray}\label{ls.10}
(\delta\tilde G_{\mu\nu})Dx^\mu Dx^\nu=2\frac{(\epsilon e_2)\dot{}}{e_2}(pDx).
\end{eqnarray}

Using these formulas variation of the action (\ref{3.3a}) under the transformations (\ref{ls.4})-(\ref{ls.6}) reads, up
to total-derivative terms
\begin{eqnarray}\label{ls.11}
\delta L=\frac12(\delta\tilde G_{\mu\nu})Dx^\mu Dx^\nu+p_\mu\delta
Dx^\mu-\left(\frac{\dot\omega^A}{e_3}\right)^.\delta\omega^A= \qquad \qquad \quad \cr \frac{(\epsilon
e_2)\dot{}}{e_2}(pDx)- \qquad \qquad \qquad \qquad \qquad \qquad \quad  \cr p_\mu \left[(\epsilon
Dx^\mu)\dot{}+\epsilon e_2\left[\omega^5\left(\frac{\dot\omega^\mu}{e_3}\right)^.-
\left(\frac{\dot\omega^5}{e_3}\right)^.\omega^\mu\right]+\frac{e_2}{e_3}\delta(\omega^5\dot\omega^\mu-\dot\omega^5\omega^\mu)\right]+
\cr \epsilon e_2p_\mu\left[\omega^5\left(\frac{\dot\omega^\mu}{e_3}\right)^.-
\left(\frac{\dot\omega^5}{e_3}\right)^.\omega^\mu\right]= \qquad \qquad \qquad \qquad  \cr \frac{(\epsilon
e_2)\dot{}}{e_2}(pDx)+\epsilon(\dot pDx)-\frac{e_2}{e_3}p_\mu\delta(\omega^5\dot\omega^\mu-\dot\omega^5\omega^\mu).
\qquad \qquad \quad
\end{eqnarray}
Compute the last term
\begin{eqnarray}\label{ls.12}
-\frac{e_2}{e_3}p_\mu\delta(\omega^5\dot\omega^\mu-\dot\omega^5\omega^\mu)= \qquad \qquad \qquad \qquad \cr
\frac{e_2}{e_3}\left(\epsilon e_2(\omega p)(\dot\omega p)+\omega^5(\epsilon e_2\omega^5p^\mu)\dot{}p_\mu-(\epsilon
e_2(\omega p))\dot{}(\omega p)-\dot\omega^5\epsilon e_2\omega^5p^2\right)= \cr \frac{e_2}{e_3}\left(\epsilon e_2(\omega
p))(\omega p)\dot{}-\epsilon e_2(\omega p)(\omega\dot p)+\omega^5\epsilon e_2\dot\omega^5p^2+(\omega^5)^2(\epsilon
e_2)\dot{}p^2+\right. \cr \left.(\omega^5)^2\epsilon e_2(\dot p p)-(\epsilon e_2)\dot{}(\omega p)^2-\epsilon e_2(\omega
p)\dot{}(\omega p)-\dot\omega^5\epsilon e_2\omega^5p^2\right)= \cr \frac{e_2}{e_3}\left((\omega^5)^2\epsilon e_2\dot
p(\eta-\frac{\omega\omega}{(\omega^5)^2})p+(\omega^5)^2(\epsilon
e_2)\dot{}p(\eta-\frac{\omega\omega}{(\omega^5)^2})p\right)= \cr -\epsilon(\dot pG\tilde GDx)-\frac{(\epsilon
e_2)\dot{}}{e_2}(pG\tilde GDx)= \qquad \qquad \qquad \cr -\epsilon(\dot pDx)-\frac{(\epsilon e_2)\dot{}}{e_2}(pDx).
\qquad \qquad \qquad \qquad \quad
\end{eqnarray}
So the last term in (\ref{ls.11}) cancel exactly the first two terms.

Let us present some other possible forms of the action (\ref{3.1}).

The action implies the kinematic constraint $(\omega^A)^2+a_4=0$
which is taken into account with help of the Lagrangian multiplier
$e_4$. According to classical mechanics \cite{arn, aadbook}, we
can solve the constraint and substitute the result back into the
action, thus obtaining
\begin{eqnarray}\label{ac.1}
L=-\frac{e_3}{2(e_2\om^5)^2}\left(\dot x^2+\frac{1}{a_4}(\dot x\om)^2\right)+ \frac{1}{e_2\om^5}(\dot
x\dot\om)-\frac{e_2}{2}a_2-\frac{e_3}{2}a_3,
\end{eqnarray}
where now $\omega^5\equiv\pm\sqrt{a_4+\omega^2}$. Now, if we omit the term $\omega^A\dot\omega_A$ in (\ref{3.1}), it
acquires the form\footnote{The actions (\ref{ac.1}) and (\ref{ac.2}) can be further simplified by rescaling
$e_2\omega^5=\tilde e_2$.}
\begin{eqnarray}\label{ac.2}
L=-\frac{e_3}{2(e_2\om^5)^2}\left(\dot x^2-\frac{1}{(\omega^A)^2}(\dot x\om)^2\right)+\frac{1}{e_2\om^5}(\dot
x\dot\om)- \cr \frac{e_2}{2}a_2-\frac{e_3}{2}a_3-\frac{e_4}{2}[(\om^A)^2+a_4]\equiv \qquad \qquad ~  \cr \frac12\tilde
G_{\mu\nu}\dot x^\mu\dot x^\nu+\frac{1}{e_2\om^5}(\dot
x\dot\om)-\frac{e_2}{2}a_2-\frac{e_3}{2}a_3-\frac{e_4}{2}[(\om^A)^2+a_4]
\end{eqnarray}
It is equivalent to (\ref{3.1}) since excluding the kinematic constraint we arrive at the expression (\ref{ac.1}). As
compared with (\ref{3.1}), the action (\ref{ac.2}) does not involve derivatives of $\omega^5$.

The kinetic part of the actions presented above contains the cross-term $(\dot x\dot\omega)$. Starting from the action
(\ref{3.3a}), we can diagonalize its kinetic part in an appropriately chosen variables. To achieve this, we introduce
the auxiliary variable $\sigma(\tau)$ and use the first-order trick \cite{aadbook} to replace
\begin{eqnarray}\label{ac.3}
-\frac{(Dx\omega)^2}{(\omega^A)^2} \quad \mbox{by} \quad \sigma^2(\omega^A)^2+2\sigma(Dx\omega).
\end{eqnarray}
Then the first term in (\ref{3.3a}) reads
\begin{eqnarray}\label{ac.4}
\frac12\tilde G_{\mu\nu}Dx^\mu Dx^\nu=-\frac{e_3}{2(e_2\om^5)^2}\left[(Dx)^2-\frac{(Dx\omega)^2}{(\omega^A)^2}\right]=
\cr -\frac{e_3}{2(e_2\om^5)^2}\left[(Dx)^2+2\sigma(Dx\omega)+\sigma^2(\omega^A)^2\right]= \cr
-\frac{e_3}{2(e_2\om^5)^2}\left[(Dx^\mu+\sigma\omega^\mu)^2-\sigma^2(\omega^5)^2\right].
\end{eqnarray}
Now the cross-terms contained in the expression $Dx^\mu+\sigma\omega^\mu$ can be diagonalized as follows
\begin{eqnarray}\label{ac.5}
Dx^\mu+\sigma\omega^\mu\equiv\dot{\tilde x}^\mu+\tilde\sigma\omega^\mu.
\end{eqnarray}
We have introduced the new variables (the variable $\tilde x^5$ will appear below)
\begin{eqnarray}\label{ac.6}
\tilde x^\mu=x^\mu-\frac{e_2\omega^5}{e_3}\omega^\mu, \cr
\tilde x^5=0-\frac{e_2\omega^5}{e_3}\omega^5, ~
\end{eqnarray}
\begin{eqnarray}\label{ac.7}
\tilde \sigma=\sigma+\left(\frac{e_2\omega^5}{e_3}\right)\dot{}+\frac{e_2\dot\omega^5}{e_3}.
\end{eqnarray}
{\it Comment.} Hamiltonian formulation of the theory with $\sigma$\,-variable implies two second-class constraints
associated with this variable. The derivative-dependent transformation (\ref{ac.7}) represents an example of conversion
of the second-class constraints, see \cite{aad5} for the details.

In the new variables, the last term in (\ref{ac.4}) reads
\begin{eqnarray}\label{ac.8}
-\sigma^2(\omega^5)^2=-\left(\tilde\sigma\omega^5-(\frac{e_2\omega^5}{e_3})\dot{}\omega^5-\frac{e_2\omega^5}{e_3}\dot\omega^5\right)^2=
\cr -\left(\tilde\sigma\omega^5+\left(-\frac{e_2(\omega^5)^2}{e_3}\right)\dot{}\right)^2=-\left(\dot{\tilde
x}^5+\tilde\sigma\omega^5\right)^2,
\end{eqnarray}
where we have replaced the variable $e_2$ by $\tilde x^5$ of Eq. (\ref{ac.6}). $\tilde x^5$ together with $\tilde
x^\mu$ form a five-dimensional space with the metric $\eta_{AB}=(-, +, +, +, -)$

Using the expressions (\ref{ac.4})-(\ref{ac.8}) in Eq. (\ref{3.3a}), the latter acquires the form
\begin{eqnarray}\label{ac.9}
L=-\frac{1}{2e_3}\left(\frac{\omega^5}{\tilde x^5}\right)^2\left(\dot{\tilde
x}^A+\tilde\sigma\omega^A\right)^2+\frac{1}{2e_3}(\dot\omega^A)^2+ \cr \frac{e_3\tilde
x^5}{2(\omega^5)^2}a_2-\frac{e_3}{2}a_3-\frac{e_4}{2}[(\om^A)^2+a_4].
\end{eqnarray}
This almost five-dimensional form of the Lagrangian has been
obtained also in \cite{aad3}. If we use the first-order trick
(\ref{ac.3}) to exclude the $\tilde\sigma$\,-variable, the
Lagrangian reads
\begin{eqnarray}\label{ac.10}
L=\frac12 G_{AB}\dot{\tilde x}^A\dot{\tilde x}^B+\frac{1}{2e_3}(\dot\omega^A)^2+\frac{e_3\tilde
x^5}{2(\omega^5)^2}a_2-\frac{e_3}{2}a_3-\frac{e_4}{2}[(\om^A)^2+a_4],
\end{eqnarray}
where the five-dimensional "metric" is
\begin{eqnarray}\label{ac.11}
G_{AB}=-\frac{1}{e_3}\left(\frac{\omega^5}{\tilde
x^5}\right)^2\left[\eta_{AB}-\frac{\omega_A\omega_B}{(\omega^C)^2}\right].
\end{eqnarray}
In contrast to $\tilde G_{\mu\nu}$, the matrix $G_{AB}$ has
null-vector $\omega^B$, $G_{AB}\omega^B=0$, hence it is not
invertible.

In the rest of this Section, we analyze the Euler-Lagrange equations which implies the action (\ref{3.3a}).  They read
\begin{eqnarray}\label{a1}
\frac{\delta S}{\delta e_2}=0 \Rightarrow \tilde
G_{\mu\nu}Dx^\mu\dot x^\nu+\frac12mc\hbar e_2=0,
\end{eqnarray}
\begin{eqnarray}\label{a2}
\frac{\delta S}{\delta e_3}=0 \Rightarrow \tilde
G_{\mu\nu}Dx^\mu(\dot
x^\nu+I^{5\nu})-\frac{1}{e_3}(\dot\omega^A)^2-a_3e_3=0,
\end{eqnarray}
\begin{eqnarray}\label{a3}
\frac{\delta S}{\delta e_4}=0 \Rightarrow (\omega^A)^2+a_4=0,
\quad ~  \mbox{then} \quad ~ \omega^A\dot\omega_A=0,
\end{eqnarray}
\begin{eqnarray}\label{a4}
\frac{\delta S}{\delta x^\mu}=0 \Rightarrow [\tilde G_{\mu\nu}Dx^\nu]^.=0,
\end{eqnarray}
\begin{eqnarray}\label{a5}
\frac{\delta S}{\delta\omega^\mu}=0 \Rightarrow
\left[\frac{\dot\omega_\mu}{e_3}-\frac{e_2}{e_3}\omega^5\tilde
G_{\mu\nu}Dx^\nu\right]\dot {}+ \qquad \cr \left[\frac{(\omega
Dx)}{(\omega^A)^2}-\frac{e_2\dot\omega^5}{e_3}\right]\tilde
G_{\mu\nu}Dx^\nu+e_4\omega_\mu=0,
\end{eqnarray}
\begin{eqnarray}\label{a6}
\frac{\delta S}{\delta\omega^5}=0 \Rightarrow
\left[\frac{\dot\omega^5}{e_3}-\frac{e_2}{e_3}\omega^\mu\tilde
G_{\mu\nu}Dx^\nu\right]\dot {}-\frac{1}{\omega^5}\tilde
G_{\mu\nu}Dx^\mu Dx^\nu+ \cr
\frac{e_3}{e_2^2\omega^5}\left[\frac{(\omega
Dx)}{(\omega^A)^2}\right]^2-\frac{e_2}{e_3}\dot\omega^\mu\tilde
G_{\mu\nu}Dx^\nu+e_4\omega^5=0. \qquad
\end{eqnarray}
The equations (\ref{a1})-(\ref{a3}) do not contain the
second-order derivative and thus represent the Lagrangian
constraints.

Equation (\ref{a4}) can be immediately integrated out
\begin{eqnarray}\label{a7}
\tilde G_{\mu\nu}Dx^\nu=p_\mu=\mbox{const}.
\end{eqnarray}
Contracting it with $G^{\alpha\mu}$, this can be written in the
form
\begin{eqnarray}\label{a8}
Dx^\mu=G^{\mu\nu}p_\nu, \quad \mbox{or} \quad \dot x^\mu=I^{5\mu}+G^{\mu\nu}p_\nu, \quad \mbox{or} \quad \dot
x^\mu=\frac{e_2}{2}\tilde J^{5\mu},
\end{eqnarray}
if we introduce the notation
\begin{eqnarray}\label{a9}
\tilde J^{5\mu}=\frac{2}{e_2}(I^{5\mu}+G^{\mu\nu}p_\nu).
\end{eqnarray}
Equation (\ref{a8}) can be compared with Eq. (\ref{3.001}). The
quantity $\tilde J^{5\mu}$ represents the Lagrangian counterpart
of angular-momentum components $J^{5\mu}$, see Eq. (\ref{1.6}).

Using (\ref{a7}), the Lagrangian constraint (\ref{a1}) reads
$(p\dot x)+\frac12 mc\hbar e_2=0$. Replacing $\dot x$ with help of
Eq. (\ref{a8}) we arrive at the classical analogy of the Dirac
equation
\begin{eqnarray}\label{a10}
p_\mu\tilde J^{5\mu}+mc\hbar=0.
\end{eqnarray}
The Lagrangian constraint (\ref{a2}) can be rewritten in two
different forms. First, using the expression $Dx=\dot x-I^5$, we
obtain
\begin{eqnarray}\label{a10.1}
\tilde G_{\mu\nu}\dot x^\mu\dot x^\nu-a_3e_3-\tilde
G_{\mu\nu}I^{5\mu}I^{5\nu}-\frac{1}{e_3}(\dot\omega^A)^2=0.
\end{eqnarray}
The constraint (\ref{a3}) implies the identity
\begin{eqnarray}\label{a11}
\tilde G_{\mu\nu}I^{5\mu}I^{5\nu}+\frac{1}{e_3}(\dot\omega^A)^2=0,
\end{eqnarray}
which can be verified by direct computations. Using it in
(\ref{a10.1}), we represent the Lagrangian constraint in the form
\begin{eqnarray}\label{a12}
\tilde G_{\mu\nu}\dot x^\mu\dot x^\nu-a_3e_3=0.
\end{eqnarray}
Using the manifest form of $\tilde G_{\mu\nu}$, we separate $\dot x^2$ as follows
\begin{eqnarray}\label{a12.1}
(\dot x^\mu)^2=-\left[\frac{(\omega\dot x)^2}{a_4}+(e_2\omega^5)^2a_3\right]<0.
\end{eqnarray}
Since $\dot x^\mu$ is the time-like four-vector, the particle's speed can not exceed the speed of light. Thus the
constraint (\ref{a2}) guarantees causal propagation of the particle.

Second, using Eqs. (\ref{a1}) and (\ref{a7}) we can exclude $x^\mu$ from (\ref{a2}). Then it reads
\begin{eqnarray}\label{a13}
(pI^5)=\frac{1}{e_3}(\dot\omega^A)^2+\frac12 mc\hbar e_2+a_3e_3.
\end{eqnarray}
Below we present it in a more transparent form, see Eq.
(\ref{a22}).

To simplify the equations (\ref{a5}) and (\ref{a6}) for the variables $\omega^A$, we observe two consequences of Eq.
(\ref{a8}).

Using (\ref{a7}), (\ref{a8}) as well as the manifest form of
$G^{\mu\nu}$, we write $\tilde G_{\mu\nu}Dx^\mu Dx^\nu= p_\mu
Dx^\mu=p_\mu
G^{\mu\nu}p_\nu=-\frac{e_2^2(\omega^5)^2}{e_3}[p^2-\frac{(\omega
p)^2}{(\omega^5)^2}]$, hence
\begin{eqnarray}\label{a14}
\tilde G_{\mu\nu}Dx^\mu Dx^\nu=-\frac{e_2^2}{e_3}[(\omega^5)^2p^2-(\omega p)^2].
\end{eqnarray}

Contracting (\ref{a7}) with $\omega^\mu$, we obtain the following expression for $(\omega Dx)$
\begin{eqnarray}\label{a15}
\frac{(\omega Dx)}{(\omega^A)^2}=\frac{e_2^2}{e_3}(\omega p).
\end{eqnarray}
These equalities together with the equation (\ref{a7}) allow us to exclude $x^\mu$ from equations (\ref{a5}) and
(\ref{a6})
\begin{eqnarray}\label{a16}
\left[\frac{\dot\omega_\mu}{e_3}-\frac{e_2}{e_3}\omega^5 p_\mu\right]\dot {}-e_2\left[\frac{\dot\omega^5}
{e_3}-\frac{e_2}{e_3}(\omega p)\right]p_\mu +e_4\omega_\mu=0,
\end{eqnarray}
\begin{eqnarray}\label{a17}
\left[\frac{\dot\omega^5}{e_3}-\frac{e_2}{e_3}(\omega p) \right]\dot
{}-e_2\left[\frac{\dot\omega_\mu}{e_3}-\frac{e_2}{e_3}\omega^5 p_\mu\right] p^\mu+e_4\omega^5=0.
\end{eqnarray}
If we introduce the notation\footnote{In the next section we
confirm that the quantities $p^\mu$ and $\pi^A$ are just the
canonical momenta of the Hamiltonian formulation.}
\begin{eqnarray}\label{a18}
\pi_\mu\equiv\frac{\dot\omega_\mu}{e_3}-\frac{e_2}{e_3}\omega^5 p_\mu, \qquad
\pi^5\equiv\frac{\dot\omega^5}{e_3}-\frac{e_2}{e_3}(\omega p),
\end{eqnarray}
these equations acquire the form
\begin{eqnarray}\label{a19}
\dot{\pi}^\mu=e_2\pi^5 p^\mu-e_4\omega^\mu,
\end{eqnarray}
\begin{eqnarray}\label{a20}
\dot{\pi}^5=e_2(\pi p)-e_4\omega^5.
\end{eqnarray}
Under the condition (\ref{a3}), the quantities $\pi^A$ obey the
relation
\begin{eqnarray}\label{a21}
\pi^A\omega_A=0,
\end{eqnarray}
which can be verified by direct computation. Besides, $e_3\pi^A\pi_A$ turns out to be proportional to $-(
pI^5)+\frac{1}{e_3}(\dot\omega^A)^2+\frac12 mc\hbar$. Taking into account (\ref{a13}), we obtain
\begin{eqnarray}\label{a22}
\pi^A\pi_A+a_3=0.
\end{eqnarray}
Thus the Lagrangian constraint (\ref{a2}) is equivalent to (\ref{a22}).

Summing up, in the notation (\ref{a9}) and (\ref{a18}) the Lagrangian equations (\ref{a1}), (\ref{a2}),
(\ref{a4})-(\ref{a6}) can be presented in an equivalent form as follows
\begin{eqnarray}\label{a24}
p_\mu\tilde J^{5\mu}+mc\hbar=0.
\end{eqnarray}
\begin{eqnarray}\label{a25}
\pi^A\pi_A+a_3=0.
\end{eqnarray}
\begin{eqnarray}\label{a26}
\dot x^\mu=\frac{e_2}{2}\tilde J^{5\mu},
\end{eqnarray}
\begin{eqnarray}\label{a27}
\dot{\pi}^\mu=e_2\pi^5 p^\mu-e_4\omega^\mu,
\end{eqnarray}
\begin{eqnarray}\label{a260}
\dot{\pi}^5=e_2(\pi p)-e_4\omega^5.
\end{eqnarray}
As it should be, they coincide with the Hamiltonian equations (\ref{3.001}), (\ref{3.0021})-(\ref{3.00310}).

\section{Canonical analysis and quantization}
Our Lagrangian action (\ref{3.3a}) does not involve derivatives of
the variables $e_a$, hence it represents an example of singular
Lagrangian theory. The Hamiltonian formulation in this case is
obtained according to the Dirac procedure for hamiltonization of a
constrained system. We do it here, with the aim to confirm that
the Lagrangian (\ref{3.3a}) and the Hamiltonian (\ref{3.0})
variational problems are actually equivalent.

For this aim, the most convenient form of the action turns out to
be those written in Eq. (\ref{3.3a}). In this form, equations for
conjugate momenta
\begin{equation}\label{h.1}
p_{\mu}=\frac{\p L}{\p \dot x^{\mu}}=\tilde G_{\mu\nu}Dx^\nu=\tilde G_{\mu\nu}(\dot x^\nu-I^{5\nu}),
\end{equation}
\begin{equation}\label{h.2}
\pi_{\mu}=\frac{\p L}{\p\dot\om^{\mu}}=\frac{\dot\omega_\mu}{e_3}-\frac{e_2}{e_3}\omega^5\tilde G_{\mu\nu}Dx^\nu=
\frac{\dot\omega_\mu}{e_3}-\frac{e_2}{e_3}\omega^5p_\mu,
\end{equation}
\begin{equation}\label{h.3}
\pi^5=\frac{\p L}{\p\dot\om_5}=\frac{\dot\omega^5}{e_3}-\frac{e_2}{e_3}\omega^\mu\tilde
G_{\mu\nu}Dx^\nu=\frac{\dot\omega^5}{e_3}-\frac{e_2}{e_3}(\omega p),
\end{equation}
can be immediately resolved with respect to velocities as follows:
\begin{equation}\label{h.4}
\dot\omega^\mu=e_3\pi^\mu+e_2\omega^5p^\mu,
\end{equation}
\begin{equation}\label{h.5}
\dot\omega^5=e_3\pi^5+e_2(\omega p),
\end{equation}
\begin{equation}\label{h.6}
\dot x^\mu=G^{\mu\nu}p_\nu+I^{5\mu}=e_2(\omega^5\pi^\mu-\omega^\mu\pi^5)=\frac{e_2}{2}J^{5\mu}.
\end{equation}
The second equality in (\ref{h.6}) follows with use of equations
(\ref{h.4}) and (\ref{h.5}). Conjugate momenta for the variables
$e_l$ turn out to be the primary constraints
\begin{equation}\label{h.6.1}
\pi_{e_l}=\frac{\p L}{\p\dot e_l}=0; \qquad l=2,3,4.
\end{equation}
Using the equalities (\ref{h.4})-(\ref{h.6.1}) we exclude
velocities from the expression $p_\mu\dot
x^\mu+\pi_A\dot\omega^A+\pi_{el}\dot e_l-L+\lambda_{el}\pi_{el}$,
and obtain the complete Hamiltonian
\begin{eqnarray}\label{h.7}
H=\frac{e_2}{2}(p_{\mu}J^{5\mu}+mc\hbar)+
\frac{e_3}{2}[(\pi^A)^2+a_3]+\frac{e_4}{2}[(\om^A)^2+a_4]+\lambda_{e_l}\pi_{e_l}.
\end{eqnarray}
As expected, it coincides with Eq. (\ref{3.01}). The procedure of
revealing the higher-stage constraints have been described in
Sect. 3. We have obtained the following chains of constraints
\begin{eqnarray}\label{h.8}
\pi_{e2}=0 ~ \Rightarrow ~ T_2\equiv p_\mu J^{5\mu}+mc\hbar=0. \qquad \qquad \qquad \qquad \qquad
\end{eqnarray}
\begin{eqnarray}\label{h.9}
\left.
\begin{array}{c}
\pi_{e3}=0,  ~ \Rightarrow ~ T_3\equiv(\pi^A)^2+a_3=0\\
\pi_{e4}=0,  ~ \Rightarrow ~ T_4\equiv(\omega^A)^2+a_4=0
\end{array}
\right\} \Rightarrow T_5\equiv\omega^A\pi_A=0, \Rightarrow \cr e_4-\frac{a_3}{a_4}e_3=0, ~ \Rightarrow ~
\lambda_{e_4}=\frac{a_3}{a_4}\lambda_{e3}. \qquad \qquad
\end{eqnarray}
The constraints $T_2$ and $\tilde T_3=T_3+\frac{a_3}{a_4}T_4$ form the first-class subset.

If we use the Dirac bracket to take into account the constraints
(\ref{h.9}), the Hamiltonian can be presented in terms of
$\varepsilon$\,-invariant variables. First, we take into account
the second-class pairs $e_4-\frac{a_3}{a_4}e_3$, $\pi_{e_4}$ and
$T_4$, $T_5$. Denoting the constraints by $K_l$, the corresponding
Dirac bracket of two phase-space functions $f$ and $g$ is
\begin{eqnarray}\label{h.10}
\{f, g\}_{1}=\{f, g\}-\{f, K_l\}\{K_l, K_m\}^{-1}\{K_m, g\}=\{f,
g\}+ \cr \frac{1}{2(\omega^A)^2}\{f,
(\omega^A)^2\}\{\omega^A\pi_A, g\}-\frac{1}{2(\omega^A)^2}\{f,
\omega^A\pi_A\}\{(\omega^A)^2, g\}+ \cr \{f,
e_4-\frac{a_3}{a_4}e_3\}\{\pi_{e4}, g\}-\{f,
\pi_{e4}\}\{e_4-\frac{a_3}{a_4}e_3, g\}, \qquad \qquad
\end{eqnarray}
where $\{,\}$ stands for the Poisson brackets. For the basic variables it reads
\begin{eqnarray}\label{h.11}
\{\om^A, \pi^B \}_{1} = \eta^{AB} - \frac{1}{(\om^A)^2}\om^A
\om^B, \quad
\{\om^A, \om^B \}_{1} = 0,
\end{eqnarray}
\begin{eqnarray}\label{h.13}
\{\pi^A, \pi^B \}_{1} = \frac{1}{(\om^A)^2}(\om^B \pi^A - \om^A
\pi^B)= -\frac{1}{2(\om^A)^2}J^{AB}.
\end{eqnarray}
Second, we impose the gauge conditions $e_3=\mbox{const}$,
$\omega^5=\mbox{const}$ for the first-class constraints $\pi_{e3}$
and $T_3$, and construct the Dirac bracket for this set of
second-class functions
\begin{eqnarray}\label{h.14}
\{f, g\}_{DB}=\{f, g\}_1+\{f, \om^5\}_{1}\Delta\{\pi^A \pi_A,
g\}_{1}-\{f, \pi^A\pi_A\}_{1}\Delta\{\om^5, g\}_{1}+ \cr \{f,
e_3\}_{1}\{\pi_{e3}, g\}_{1}-\{f, \pi_{e3}\}_{1}\{e_3, g\}_{1},
\qquad \qquad
\end{eqnarray}
where
\begin{eqnarray}\label{h.15}
\Delta=\frac{(\om^A)^2}{2[(\om^A)^2\pi^5-(\om^A \pi_A)\om^5]}.
\end{eqnarray}
It implies
\begin{eqnarray}\label{h.15.3}
\{\om^A, \pi^B
\}_{DB}=G^{AB}-\frac{J^{AC}\omega_CG^{5B}}{J^{5C}\omega_C}, \quad
\{\om^A, \om^B\}_{DB} = 0.
\end{eqnarray}
\begin{eqnarray}\label{h.15.2}
\{\pi^A, \pi^B
\}_{DB}=-\frac{J^{AB}}{2(\omega^C)^2}+\frac{G^{5A}J^{BC}\pi_C-(A\leftrightarrow
B)}{J^{5C}\omega_C}.
\end{eqnarray}
We have denoted (see also Eq. (\ref{ac.11}))
\begin{eqnarray}\label{h.15.4}
G^{AB}=\left[\eta^{AB}-\frac{\om^A\om^B}{(\om^C)^2}\right].
\end{eqnarray}
These formulas acquire more simple form on the constraint surface
\begin{eqnarray}\label{h.16.3}
\{\om^A, \pi^B \}_{DB}=G^{AB}-\frac{\pi^AG^{5B}}{\pi^5}, \quad
\{\om^A, \om^B \}_{DB}=0.
\end{eqnarray}
\begin{eqnarray}\label{h.16.2}
\{\pi^A, \pi^B \}_{DB}=\frac{J^{AB}}{2a_4}-\frac{a_3}{a_4\pi^5}\eta^{5[A}\omega^{B]}.
\end{eqnarray}

The Dirac bracket of the basic variables is deformed as compared with the Poisson one. In contrast, brackets of
$J^{AB}$ keep their initial form. Indeed, the constraints $T_a$ are $SO(2, 3)$\,-invariants, $\{T_a, J^{AB}\}=0$. On
this reason, if we compute the Dirac bracket of the spin-tensor components, all its extra-terms vanish, $\{J^{AB},
J^{CD}\}_{DB}$ $=$ $\{J^{AB}, J^{CD}\}$. Thus the transition from Poisson to the Dirac bracket does not modify the
initial spin-tensor algebra (\ref{1.4.1}).

By construction, the Dirac bracket of a constraint with any
function vanishes identically. So, when we are dealing with the
Dirac bracket, the constraints can be used in all the computations
as strong equations. In particular, we can omit them in the
expression (\ref{h.7}). It gives the $\varepsilon$\,-invariant
Hamiltonian
\begin{eqnarray}\label{h.7.1}
H_1=\frac{e_2}{2}(p_{\mu}J^{5\mu}+mc\hbar)+\lambda_{e_2}\pi_{e_2}.
\end{eqnarray}
The Hamiltonian equations (\ref{3.001})-(\ref{3.002}),
(\ref{3b.6}), (\ref{3b.61}) can now be obtained according the rule
$\dot A=\{A, H_1\}_{DB}$ instead of $\dot A=\{A, H\}$.

As we have shown in Sect. 2, it is consistent to describe
spin-sector of the model by the $\varepsilon$\,-invariant
variables $J^{AB}$ instead of the initial coordinates $\omega^A$,
$\pi^B$. Canonical quantization of the model is achieved replacing
the variables $x^\mu$, $p^\nu$ and $J^{AB}$ by operators which
obey the rule $[ ~ , ~ ]=i\hbar\{ ~ , ~ \}$, where the
nonvanishing classical brackets are
\begin{eqnarray}\label{h.17}
\{x^\mu, p^\nu\}=\eta^{\mu\nu}
\end{eqnarray}
\begin{eqnarray}\label{h.18}
\{J^{AB}, J^{CD}\}=2(\eta^{AC} J^{BD}-\eta^{AD}J^{BC}-\eta^{BC}J^{AD}+\eta^{BD}J^{AC}).
\end{eqnarray}
The operators
\begin{eqnarray}\label{1.999}
\hat x^\mu=x^\mu, \qquad \hat p_\mu=-i\hbar\partial_\mu, \cr \hat J^{5\mu}=\hbar\Gamma^\mu, \qquad \hat
J^{\mu\nu}=\hbar\Gamma^{\mu\nu},
\end{eqnarray}
obey the necessary commutators. They act on the space of Dirac
spinors $\Psi(x^\mu)$. The only constraint which has not been yet
taken into account is $T_2=p_\mu J^{5\mu}+mc\hbar=0$. Since this
is the first-class function, it is consistent to impose the
corresponding operator on a state vector. This gives the Dirac
equation, $(\Gamma^\mu\hat p_\mu+mc)\Psi=0$.

The classical equations of motion (\ref{3b.7}) and (\ref{3b.71}) imply that the center-of-charge coordinate $x^i$
experiences  a complicated trembling motion in the theory without interaction (see also the next section). Besides the
operator $\hat x$, some other versions for the position operator in the Dirac theory have been suggested and discussed
in the literature \cite{pr, 7, nw}. In our model we can construct the following "center-of-mass" variable
\begin{eqnarray}\label{2.13.00}
\tilde x^\mu=x^\mu+\frac{1}{2p^2}J^{\mu\nu}p_\nu.
\end{eqnarray}
It obeys the equation
\begin{equation}\label{4.34}
\dot {\tilde x}^{\mu}=\tilde e p^{\mu}; \quad \mbox{where} \quad
\tilde e\equiv-\frac{e_2mc\hbar}{2p^2},
\end{equation}
Note also that $p^\mu$ represents the mechanical momentum of the $\tilde x^\mu$\,-particle. The
reparametrization-invariant quantity $\tilde x^i(t)$ moves along the straight line, $\frac{d\tilde
x^i}{dt}=\frac{cp^i}{p^0}$. We propose the variable $\tilde x^\mu$ as the Lorentz-covariant analog of the
Foldy-Wouthuysen operator.

Using components of the spin-tensor, we can construct the
Pauli-Lubanski vector $S^\mu=\epsilon^{\mu\nu\alpha\beta}p_\nu
J_{\alpha\beta}$. It has no precession in the free theory, $\dot
S^\mu=0$, and corresponds to the Bargmann-Michel-Telegdi
spin-vector \cite{bmt}.

In the center-of-charge instantaneous rest frame, $J^{50}$ $=$
$\mbox{const}$, $J^{5i}$ $=$ $0$, it reduces to $S^\mu=(0,
S^i=\epsilon^{ijk}p_jJ_{0k})$. The only $J_{0k}$\,-part of the
angular-momentum tensor survives in this frame.

In the center of mass frame, $p^{\mu}=(p^0, 0, 0, 0)$, $\vec S$ is
proportional to the three-dimensional rotation generator,
$S^{\mu}=(0, \frac{1}{2}p^0\epsilon^{ijk}J_{jk})$.

The equation (\ref{2.13.00}) represents the phase-space transformation which is not the canonical one.  As a
consequence, the theory became non-commutative. Computing the Poisson brackets, we obtain
\begin{eqnarray}
\{\ti x^\mu, \ti x^\nu\}=-\frac{1}{2p^2}\left(J^{\mu \nu}+\frac{1}{p^2}p^{[\mu}J^{\nu ]\al}p_\al\right), \quad \{\ti
x^\mu, p^\nu\}=\eta^{\mu \nu},
\end{eqnarray}
\begin{eqnarray}
\{\ti x^\mu, J^{5\nu}\}=\frac{1}{p^2}(\eta^{\mu \nu}J^{5\al}p_\al-J^{5\mu}p^\nu).
\end{eqnarray}
\begin{eqnarray}
\{\ti x^\mu, J^{\al \be}\}=\frac{1}{p^2}(J^{\mu[\al}p^{\be]}-\eta^{\mu[\al}J^{\be]\gamma}p_\gamma),
\end{eqnarray}
We also present the brackets with the Pauli-Lubanski vector
\begin{eqnarray}
\{S^\mu, S^\nu \}=2p^2J^{\mu \nu}-2p^{[\mu}J^{\nu ]\al}p_\al.
\end{eqnarray}
\begin{eqnarray}
\{\ti x^\mu, S^\nu \} =-\frac{1}{2}\ep^{\mu\nu\rho\sigma}J_{\rho\sigma}-\frac{1}{p^2}\ep^{\mu\nu\gamma \sigma}p_\gamma
J_{\sigma}{}^{\al}p_\al,
\end{eqnarray}
\begin{eqnarray}
\{S^\mu, J^{5 \al} \} = 2  \ep^{\mu \al \sigma \nu}J^5{}_\sigma p_\nu, \quad \{S^\mu, J^{\al \be} \} = 2 p_\nu \ep^{\mu
\nu \sigma [ \al}J^{\be]}{}_\sigma.
\end{eqnarray}
We point out that the second term in Eq. (\ref{2.13.00}) has the structure typical for non-commutative extensions of
the usual mechanics, see \cite{aad7, das, yan, amo}.

\section{Solution to the classical equations of motion}
We are interested in to solve equations of motion for the gauge-invariant variables $x^i(t)$, $p^{\mu}(t)$ and
$J^{AB}(t)$, they are\footnote{Solution to equations of motion for the initial variables $\omega^A$ and $\pi^B$ as well
as the subsequent construction on this base the physical variables $x^i(t)$, and $J^{AB}(t)$ are given in the
Appendix.}
\begin{eqnarray}\label{4.3}
\frac{dx^i}{dt}=c\frac{J^{5i}}{J^{50}}, \qquad \quad \\
\label{4.32}
\frac{dJ^{5\mu}}{dt}=2cp_{\nu}\frac{J^{\nu \mu}}{J^{50}}, \qquad \\
\label{4.31} \frac{dJ^{\mu\nu}}{dt} =2c\frac{p^{\mu}J^{5\nu}-p^{\nu}J^{5\mu}}{J^{50}},
\end{eqnarray}
where $p^\mu=\mbox{const}$. Besides, the spin-tensor $J^{AB}$ and the time-like vector $J^{5\mu}$ obey the constraints
\begin{eqnarray}\label{4.3a}
p_\mu J^{5\mu}+mc\hbar=0,
\end{eqnarray}
\begin{eqnarray}\label{4.3b}
J^{AB}J_{AB}=8a_3a_4.
\end{eqnarray}
The differential equations written above are not polynomial. To
avoid this difficulty, we remind that they were obtained from the
equations for coordinates presented in parametric form
\begin{eqnarray}\label{gi.1}
\dot x^{\mu} = \frac{e_2}{2}J^{5 \mu}, \qquad \qquad \\
\label{gi.3} \dot J^{5 \mu} = e_2p_{\nu}J^{\nu \mu}, \qquad \quad \\
\label{gi.2} \dot J^{\mu \nu} = e_2(p^{\mu}J^{5 \nu}- p^{\nu}J^{5 \mu}),
\end{eqnarray}
eliminating the ambiguity due to $e_2(\tau)$. The latter equations are polynomial, so their analysis represents more
simple task. In the process, we can conventionally fix $e_2(\tau)$, since according to (\ref{4.3})-(\ref{4.3b}), the
physical coordinates we are interested in do not depend on a particular choice of $e_2$. We take $e_2=\mbox{const}$.
After integrating the equations (\ref{gi.1})-(\ref{gi.2}), we exclude $\tau$ from the resulting expressions thus
obtaining the general solution to (\ref{4.3})-(\ref{4.31}).

Let us start from Eq. (\ref{gi.3}). We compute its derivative and
use the equations (\ref{gi.2}) and (\ref{4.3a}), it gives the
closed equation for $J^{5\mu}$
\begin{eqnarray}\label{e.1}
\ddot J^{5\mu}-e^2_2p^2J^{5\mu}=e^2_2 mc\hbar p^{\mu},
\end{eqnarray}
whose solution depends on the sign of the constant $p^2$. We
discuss the two possibilities separately. \par \noindent
\subsection{Spinning-particle with $p^2<0$, helical motions}
\par \noindent
In this case, the general solution to Eq. (\ref{e.1}) is given by
\begin{eqnarray}\label{e.2}
J^{5\mu}=\frac{mc\hbar}{|p|^2}p^{\mu}+A^{\mu}\cos(\om\tau)+B^{\mu}\sin(\om\tau),
\end{eqnarray}
Where $|p|=\sqrt{-p^2}$, $\om=e_2|p|$, and $A^\mu$, $B^\mu$ are
the integration constants. According to Eq. (\ref{4.3a}), they
obey the restrictions
\begin{eqnarray}\label{e.8}
p_{\mu}A^{\mu} = 0, \qquad  p_{\mu}B^{\mu} = 0.
\end{eqnarray}
We substitute Eq. (\ref{e.2}) into (\ref{gi.1}). It gives closed
equation for $x^\mu$ which can be integrated
\begin{eqnarray}\label{e.3}
x^{\mu}(\tau)=X^{\mu}+e_2\frac{mc \hbar}{2|p|^2}p^{\mu}\tau+\frac{1}{2|p|}A^{\mu}\sin(\om\tau)-
\frac{1}{2|p|}B^{\mu}\cos(\om\tau).
\end{eqnarray}
By construction, only the monotonic functions $x^0(\tau)$ are physically admissible. This implies
\begin{eqnarray}\label{e.3.1}
A^0=B^0=0.
\end{eqnarray}
We also fix the initial instant to be zero, $X^0=0$. Hence
\begin{eqnarray}\label{e.3.3}
x^0=ct=\frac{e_2mc\hbar}{2|p|^2}p^0\tau \Rightarrow \tau=\frac{2 |p|^2}{e_2m\hbar p^0}t.
\end{eqnarray}
Similarly, if we substitute (\ref{e.2}) into the equation $\dot
J^{0i} = e_2(p^{0}J^{5 i}- p^{i}J^{5 0})$, it can be integrated as
well
\begin{eqnarray}\label{e.4}
J^{0i}=\Sigma^{0i}+\frac{1}{|p|}p^{0}A^{i}\sin(\om\tau)-\frac{1}{|p|}p^{0}B^{i}\cos(\om\tau),
\end{eqnarray}
$\Sigma^{0i}$ is the integration constant. As we know, the
remaining $J$'s are not independent, and can be computed according
the equation
\begin{eqnarray}\label{jij}
J^{ij}=(J^{50})^{-1}(J^{5i}J^{0j}-J^{5j}J^{0i}).
\end{eqnarray}
It reads
\begin{eqnarray}\label{e.4.1}
J^{ij}= \frac{p^{[i}\Sigma^{0j]}}{p^0}+\frac{|p|B^{[i} A^{j]}}{mc\hbar}+\left[\frac{1}{|p|}p^{[i}A^{j]}\right.
+\left.\frac{|p|^2}{mc\hbar p^0} B^{[i}\Sigma^{0j]}\right]\sin(\om\tau) \cr +\left[-\frac{1}{|p|}p^{[i}
B^{j]}+\frac{|p|^2}{mc\hbar p^0} A^{[i}\Sigma^{0j]}\right]\cos(\om\tau).
\end{eqnarray}

We have started our computations from the equation (\ref{e.1})
which is a consequence of (\ref{gi.3}). To select the subset of
solutions which obeys the initial system
(\ref{gi.1})-(\ref{gi.2}), we substitute (\ref{e.2}), (\ref{e.4})
and (\ref{e.4.1}) into the equation (\ref{gi.3}). It is satisfied
if
\begin{eqnarray}\label{e.4.2}
\Sigma^{0i}=0.
\end{eqnarray}

The next step is to satisfy the constraints (\ref{4.3a}),
(\ref{4.3b}). The first one have been already taken into account,
it implies (\ref{e.8}). The constraint (\ref{4.3b}) leads to the
following restriction:
\begin{eqnarray}\label{e.11}
\left(\frac{mc\hbar}{|p|}\right)^2-(\vec A^2+\vec B^2)+\left(
\frac{|p|}{mc\hbar}\right)^2(\vec A\times\vec B)^2=4a_3 a_4.
\end{eqnarray}
Besides, $J^{5\mu}$ turns out to be the time-like vector if
\begin{eqnarray}\label{e.10}
[\vec A\cos(\om\tau)+\vec B\sin
(\om\tau)]^2<\left(\frac{mc\hbar}{|p|}\right)^2, \quad \mbox{for
any} \quad \tau.
\end{eqnarray}
It implies
\begin{eqnarray}\label{e.10.1}
\vec A^2<\left(\frac{mc\hbar}{|p|}\right)^2, \qquad \vec B^2<\left(\frac{mc\hbar}{|p|}\right)^2.
\end{eqnarray}

The last step is to exclude the parameter $\tau$ from the
expressions obtained. Using Eq. (\ref{e.3.3}), we obtain the
general solution to the equations (\ref{4.3})-(\ref{4.31})
\begin{eqnarray}\label{res1}
x^i(t)=X^i+c\frac{p^i}{p^0}t+\frac{1}{2|p|}(A^i\sin(\ti\om t)-B^i\cos(\ti\om t)), \qquad \\
\label{res1.3} J^{0i}=\frac{p^{0}}{|p|}(A^{i}\sin(\ti\om\tau)-
B^{i}\cos(\ti\om\tau)), \qquad \qquad \qquad ~ \qquad \\
\label{res1.2} J^{5i}(t)=\frac{mc\hbar}{|p|^2}p^i+A^i\cos(\ti\om
t)+
B^i\sin(\ti\om t), \qquad \qquad \qquad \\
\label{res1.2.1} J^{50}(t)=\frac{mc\hbar p^0}{|p|^2},
\qquad \qquad \qquad \qquad \qquad \qquad ~  \qquad \qquad \\
\label{res1.4} J^{ij}(t)=
\frac{|p|}{mc\hbar}B^{[i}A^{j]}+\frac{1}{|p|}p^{[i}A^{j]}\sin(\ti
\om t)-\frac{1}{|p|}p^{[i}B^{j]}\cos(\ti\om t),
\end{eqnarray}
where the frequency is
\begin{eqnarray}\label{res1.6}
\tilde\omega=\frac{2|p|^3}{m \hbar p^0},
\end{eqnarray}
and the integration constants obey the restrictions (\ref{e.11}), (\ref{e.10}) and
\begin{eqnarray}\label{res2}
\vec p\vec A=0, \quad \vec p\vec B=0.
\end{eqnarray}

Let us discuss dynamics of the coordinate $x^i(t)$.

Using the equations (\ref{res2}) and (\ref{e.10}), we confirm once again its causal motion
\begin{eqnarray}\label{e.110}
\left(\frac{d \vec x}{dt}\right)^2=c^2\frac{\vec p^2}{(p^0)^2}+
\frac{|p|^4}{(m \hbar p^0)^2}(\vec A \cos (\ti \om t)+\vec
B\sin(\ti\om t))^2< \cr c^2 \frac{\vec p^2}{(p^0)^2} +
\frac{|p|^4}{(m\hbar p^0)^2}\frac{(mc \hbar)^2}{|p|^2}=c^2. \qquad
\qquad \qquad
\end{eqnarray}
The curve (\ref{res1}) is a helix which can be considered as a superposition of the rectilinear motion
\begin{eqnarray}
\tilde x^i(t)=X^i+c\frac{p^i}{p^0}t,
\end{eqnarray}
and the oscillatory motion
\begin{eqnarray}\label{z.1}
z^i(t)=\frac{1}{2|p|}(A^i\sin(\ti\om t)-B^i\cos(\ti\om t)),
\end{eqnarray}
the latter is the classical-mechanical analog of
\textit{Zitterbewegung}.

Both the conjugate momentum and the Dirac equation acquire certain interpretation in this picture. According to Eq.
(\ref{res2}), the \textit{Zitterbewegung} oscillations occur on the plane perpendicular to $\vec p$. This is the Dirac
equation that dictates the perpendicularity.

On this plane, the trajectory is an ellipse. To see this, we take
the coordinate system with the origin at $\tilde x^i$ and with the
axis $x_1$ and $x_2$ on the plane of the vectors $\vec A$ and
$\vec B$ in such a way, that the $x_1$\,-axis has the direction of
the vector $\vec B$. Then $\vec B=(B^1, 0)$ and $\vec A=(A^1,
A^2)$. In this coordinate system the parametric equations of the
\textit{Zitterbewegung} reads
\begin{eqnarray}\label{e.12}
x_1=\frac{1}{2|p|}(A^1\sin(\ti\om t)-B^1\cos(\ti\om t)), \cr
x_2=\frac{1}{2|p|}A^2\sin(\ti\om t). \qquad \qquad \qquad ~ ~
\end{eqnarray}
To obtain its trajectory, we ask whether the points of the curve
(\ref{e.12}) can be identified with those of a second-order line,
$C_{ij}x_ix_j=1$. Short computation shows that the line is given
by
\begin{eqnarray}\label{e.13}
C=\frac{4|p|^2}{(B^1)^2}\left(
\begin{array}{cc}
1 & -\frac{A^1}{A^2} \\
-\frac{A^1}{A^2} & \frac{(A^1)^2+(B^1)^2}{(A^2)^2}
\end{array}
\right).
\end{eqnarray}
Since $\det C=\left(\frac{4|p|^2}{A^2B^1}\right)^2>0$, the line
represents an ellipse. Denoting its semi-axis as $a$ and $b$, we
can write
\begin{eqnarray}\label{e.14}
\frac{1}{a^2}+\frac{1}{b^2}=\mbox{Tr} ~ C, \quad
\frac{1}{a^2b^2}=\det C, \quad \mbox{then} \quad
a^2+b^2=\frac{\mbox{Tr} ~ C}{\det C}.
\end{eqnarray}
The last equation together with (\ref{e.10.1}) allow us to
estimate the size of the ellipse as follows:
\begin{eqnarray}\label{e.15}
a^2+b^2=\frac{\vec A^2+\vec B^2}{4|p|^2}<\frac12\left(\frac{mc\hbar}{|p|^2}\right)^2.
\end{eqnarray}

As we have seen above, canonical quantization of our model in the
coordinates $x$, $p$ and $J$ leads to the Dirac equation. Now we
look for the coordinates which may have a reasonable
interpretation in the classical theory.

Let us compute the total number of physical degrees of freedom in
the theory. Omitting the auxiliary variables and the corresponding
constraints, we have $18$ phase-space variables $x^\mu$, $p_\mu$,
$\omega^A$, $\pi_A$ subject to the constraints (\ref{1.7}),
(\ref{1.71}), (\ref{1.8}). Taking into account that each
second-class constraint rules out one variable, whereas each
first-class constraint rules out two variables, the number of
physical degrees of freedom is $18-(2+2\times 2)=12$. Note that
this is equal to the number of degrees of freedom of the two-body
problem.

Further, we note that the \textit{Zitterbewegung} (\ref{z.1}) coincides  with the evolution of the coordinate
$\frac{J^{0i}}{2p^0}$ of the inner spin-space, see Eq. (\ref{res1.3}). Let us choose
\begin{eqnarray}\label{e.16}
\tilde x^i(t)= x^i-\frac{J^{0i}}{2p^0}, \qquad \tilde
p^i=\frac{p^i}{p^0};
\end{eqnarray}
\begin{eqnarray}\label{e.17}
J^i=\frac{J^{0i}}{2p^0}, \qquad
V^i=\frac{J^{5i}}{J^{50}}-\frac{p^i}{p^0}.
\end{eqnarray}
as the new coordinates in the problem\footnote{We stress that we
have performed a transformation on the phase-space.}. The spatial
coordinates obey the equations
\begin{eqnarray}\label{e.18}
\frac{d\tilde x^i}{dt}=c\tilde p^i, \qquad \frac{dJ^i}{dt}=cV^i.
\end{eqnarray}
and, according to Eqs. (\ref{res1})-(\ref{res1.2.1}), the general
solution is
\begin{eqnarray}\label{e.19}
\tilde x^i=X^i+c\frac{p^i}{p^0}t,
\end{eqnarray}
\begin{eqnarray}\label{e.20}
J^i(t)=\frac{1}{2|p|}(A^i\sin(\ti\om t)-B^i\cos(\ti\om t)).
\end{eqnarray}
As we have seen, the trajectory of $J^i$ is an ellipse on the plane perpendicular to $\vec p$. Thus, behavior of the
coordinates $\tilde x^i$ and $J^i$ is similar to those of the center-of-mass and the relative position of a two-body
system subjected to a central field.

The Dirac electron obeys the mass-shell condition $(p^\mu)^2=-m^2c^2$. Let us consider the subset of solutions in our
model with this value of $p^\mu$. Then for the two-particle system with a slowly moving center-of-mass $\tilde x^i$, we
have $p^0\sim|p|=mc$. Then the frequency (\ref{res1.6}) of the relative position $J^i$ approaches to the Compton
frequency, $\tilde\omega\sim\frac{2mc^2}{\hbar}$. Besides, the size of elliptic orbit turns out to be of the order of
de Broglie wave-length, $\sqrt{a^2+b^2}<\frac{\hbar}{\sqrt{2}mc}$.

It would be interesting to quantize the model in the coordinates
(\ref{e.16}), (\ref{e.17}) and to compare the results with those
of Dirac theory in the Foldy-Wouthuysen representation.

\subsection{Spinning-particle with $p^2>0$, hyperbolic motions} \par \noindent
The general solution to (\ref{e.1}) is now given by
\begin{eqnarray}\label{e.24}
J^{5 \mu}(\tau) = -\frac{mc \hbar}{|p|^2}p^{\mu}+
A^{\mu}e^{\om \tau} + B^{\mu}e^{-\om \tau},
\end{eqnarray}
where $|p|=\sqrt{p^2}$, $\om=e_2|p|$. The constraint (\ref{4.3a})
implies $p_{\mu}A^{\mu}=0$, $p_{\mu}B^{\mu}=0$. With this
$J^{5\mu}(\tau)$ at hands, we immediately integrate equations for
$x^{\mu}$ and $J^{\mu \nu}$
\begin{eqnarray}\label{e.25}
x^{\mu}(\tau)=X^{\mu}-\frac{e_2mc\hbar}{2|p|^2}p^{\mu}\tau+\frac{1}{2|p|}A^{\mu}e^{\om\tau}-\cr -
\frac{1}{2|p|}B^{\mu}e^{-\om\tau},
\\ \label{e.26}
J^{\mu\nu}(\tau)=\Sigma^{\mu\nu}+\frac{1}{|p|}(p^{\mu}A^{\nu}-p^{\nu}A^{\mu})e^{\om \tau}-\cr -\frac{1}{|p|}
(p^{\mu}B^{\nu}-p^{\nu}B^{\mu})e^{-\om\tau},
\end{eqnarray}
where $A^{\mu}$, $B^{\mu}$, $X^{\mu}$ and $\Sigma^{0i}$ are the
integration constants. The components $J^{ij}$ have been found
with help of (\ref{jij}), it implies the following expression for
$\Sigma^{ij}$ in terms of the integration constants
\begin{eqnarray}\label{e.25.1}
\Sigma^{ij} = \frac{2|p|}{mc\hbar}(A^iB^j - A^j B^i).
\end{eqnarray}

We have started our computations from the equation (\ref{e.1}) which is a consequence of (\ref{gi.3}). To select the
subset of solutions which obeys the initial system (\ref{gi.1})-(\ref{gi.2}), we substitute (\ref{e.24})-(\ref{e.25.1})
into the equation (\ref{gi.3}). This determines $\Sigma^{0i}$ in terms of other integration constants
\begin{eqnarray}\label{e.25.2}
\Sigma^{0i} = \frac{2|p|}{mc\hbar}(A^0B^i - B^0 A^i),
\end{eqnarray}
as well as implies the restriction
\begin{equation}\label{e.27}
A^0(\vec p\vec B)=B^0(\vec p\vec A).
\end{equation}
The equations (\ref{e.25.1}) and (\ref{e.25.2}) can be unified into the expression
\begin{eqnarray}\label{e.25.3}
\Sigma^{\mu\nu} = \frac{2|p|}{mc\hbar}A^{[\mu}B^{\nu]}.
\end{eqnarray}

The next step is to satisfy the constraints (\ref{4.3a}),
(\ref{4.3b}). They imply the following restrictions on the
integration constants
\begin{eqnarray}\label{e.8.00}
p_{\mu}A^{\mu} = 0, \qquad p_{\mu}B^{\mu} = 0, \quad \mbox{then} \quad p_\mu\Sigma^{\mu\nu}=0.
\end{eqnarray}
\begin{eqnarray}\label{e.25.4}
\left(\frac{mc\hbar}{2|p|}\right)^2+AB-\left(\frac{|p|}{mc\hbar}\right)^2[(\vec A\times\vec B)^2-(A^0\vec B-B^0 \vec
A)^2]=a_3a_4 .
\end{eqnarray}
In obtaining the last equation we have used the restrictions (\ref{e.8.00}). Besides, $J^{5\mu}$ turns out to be the
time-like vector if
\begin{eqnarray}\label{e.25.5}
\left(\frac{mc\hbar}{|p|}\right)^2+A^2e^{2\om\tau}+2AB +B^2e^{-2\om\tau}<0, \quad \mbox{for any $\tau$}.
\end{eqnarray}
Since $e^{2\om\tau}$ increases, this inequality implies, in particular
\begin{eqnarray}\label{e.28}
A_\mu A^\mu<0, \qquad B_\mu B^\mu<0.
\end{eqnarray}
The solution (\ref{e.25}) describes a self-accelerated motion. For the present case, we can not exclude the parameter
$\tau$ analytically. For sufficiently big values of $\tau$, we can neglect all the terms in the expression for $x^\mu$
as compared with the third term, then
\begin{eqnarray}
x^0(\tau)\approx\frac{1}{2|p|}A^0 e^{\om\tau} \Rightarrow e^{\om \tau}=\frac{2|p|c}{A^0}t.
\end{eqnarray}
It gives the following asymptotic for $x^i(t)$
\begin{eqnarray}
x^i(t)\approx X^i+c\frac{A^i}{A^0}t, \quad \mbox{when} \quad t\rightarrow +\infty.
\end{eqnarray}
Since $A^\mu$ is the time-like vector, see Eq. (\ref{e.28}), we have $|\vec v|<c$. The hyperbolic motions are presented
also in the Frenkel theory of an electron, see \cite{obu}.

\section{Conclusion}
In this paper we have constructed the non-Grassmann mechanical model which implies the Dirac equation. To formulate a
variational problem, we have identified the spin variables with coordinates of the base of a seven-dimensional fiber
bundle embedded as a surface into ten-dimensional phase space equipped with the canonical Poisson bracket $\{\om^A,
\pi^B\}=\eta^{AB}$ (the configuration space $\omega^A$ turns out to be the anti-de Sitter space,
$\omega^A\omega_A+a_4=0$). By construction, the values of Casimir operators of $SO(2, 3)$\,-group are fixed within the
classical theory (in the dynamical realization this is guaranteed by the first-class constraint $T_3=0$, see Eqs.
(\ref{1.7}) and (\ref{1.71.3})).

The surface can be covered by the coordinates $J^{5\mu}, J^{0i}, \omega^5$, adjusted with the structure of fiber
bundle, then $J^{5\mu}, J^{0i}$ represents coordinates of the base, while $\omega^5$ parameterizes the fiber. In
accordance with this, we can discard $\omega^5$ (in the dynamical realization $\omega^5$ is affected by the local
transformations, hence it is not observable quantity). In the result, only the coordinates of the base,  $J^{5 \mu},
J^{0i}$, are relevant to description of spin. Canonical quantization of the coordinates produces both $\Gamma^\mu$ and
$\Gamma^{\mu\nu}$\,-matrices, see Eqs. (\ref{1.71.5})-(\ref{1.9}).

It would be interesting to describe the underlying geometry of the fiber bundle (as it has been mentioned above, for
the non-relativistic spin this is $SO(3)$ fiber bundle).

Besides the geometric constraints (\ref{1.7}) and (\ref{1.71}), which determines the spin surface, our model implies
the dynamical first-class constraint (\ref{1.8}). Being imposed on a state-vector, it implies the Dirac equation.
Although there is no the mass-shell constraint $p^2+ m^2c^2=0$ in our model, our particle's speed cannot exceed the
speed of light. This is due to the geometric constraints which imply the time-like character of the four-vector
$J^{5\mu}$, see Eq. (\ref{k4}). In turn, this implies the causal dynamics of the coordinate $x^\mu$, see Eq.
(\ref{3.001}).

Analyzing the general solution (\ref{res1})-(\ref{res2.3}) to the classical equations of motion
(\ref{4.3})-(\ref{4.3b}), we have constructed the coordinates (\ref{e.16}), (\ref{e.17}) which strongly resemble the
two-body problem. If we assume space-time interpretation of the coordinates $\tilde x^i$ and $J^i$, they can be
identified with the center-of-mass and the relative position of a two-body system subjected to a central field. The
Dirac equation dictates the perpendicularity of the {\it Zitterbewegung}-plane to the direction of the center-of-mass
motion.

Dynamics of the subset of solutions with $(p^\mu)^2=-m^2c^2$ is in correspondence with the dynamics of mean values of
the corresponding operators in the Dirac theory. In particular, for the two-particle system with a slowly moving
center-of-mass $\tilde x^i$, the frequency (\ref{res1.6}) of the relative position $J^i$ approaches to the Compton
frequency, $\tilde\omega\sim\frac{2mc^2}{\hbar}$. Besides, the size of elliptic orbit turns out to be of the order of
de Broglie wave-length, $\sqrt{a^2+b^2}\sim\frac{\hbar}{mc}$.

Our model shows the same undesirable properties as those of Dirac equation in the RQM interpretation. We finish with a
brief comment on a modification which solves the problems. We recall that the Dirac equation (\ref{1.1}) implies the
Klein-Gordon one. In contrast, in classical mechanics the corresponding constraint (\ref{1.8}) does not imply the
mass-shell constraint $p^2+m^2c^2=0$. So, the model presented here is not yet in complete correspondence with the Dirac
theory. The semiclassical model that produces both constraints has been discussed in the recent work \cite{aad1}. The
extra first-class constraint implies that we are dealing with a theory with one more local symmetry, with the
constraint being a generator of the symmetry \cite{gt, aadbook, aad4, abr}. This leads to a completely different
picture of the classical dynamics. The variable $x^\mu$ is not inert under the extra symmetry. Being gauge
non-invariant, $x^\mu$ turns out to be an unobservable quantity. The variable $\tilde x^\mu$ of Eq. (\ref{2.13.00}) is
gauge invariant and should be identified with the position of the particle. Because $p^\mu$ is a mechanical momentum
for $\tilde x^\mu$, the particle's speed cannot exceed the speed of light. In the absence of interaction it moves along
the straight line. Hence the modified model is free of {\it Zitterbewegung}.

\section*{Acknowledgments}
A. A. D. and P. S. C. acknowledge financial support from the Brazilian foundation FAPEMIG. B. F. R. would like to thank
the Brazilian foundation FAPEAM - Programa RH Interioriza\c{c}\~ao - for financial support. G. P. Z. C. acknowledges
financial support from the Brazilian foundation CNPq.

\section*{Appendix}
\setcounter{equation}{0}
\def\theequation{A.\arabic{equation}}
In Section 6 we have solved equations of motion for the variables $J^{AB}(\tau)$ and $x^{\mu}(\tau)$. Here we present
solution to equations of motion for the initial variables $\om^A$, $\pi^B$ and $x^\mu$. We shall restrict ourselves to
the case $p^2 < 0$. We keep the notation $p^2 \equiv - |p|^2$. As discussed before, the equations for $J^{AB}(t)$ and
$x^i(t)$ do not depend on the auxiliary variables $e_{2,3}$. So, we are free to select them as we want. We take
$e_{2,3} = 1$ and define $a \equiv \frac{a_3}{a_4}$. The equations of motion for $\om^A$ and $\pi^B$ then read
\begin{eqnarray}\label{a.1}
\dot \om^{\mu} = \pi^{\mu}+ \om^5 p^{\mu}, \quad
\dot \pi^{\mu} = -a \om^{\mu} + \pi^5 p^{\mu}; \\
\label{a.2} \dot \om^5 = \pi^5 + p\om, \quad \dot \pi^5 = - a \om^5 + p \pi. \quad \,\,\,
\end{eqnarray}
Let us discuss one possible way to decouple the system above. The contraction of the equations for $\om^{\mu}$ and
$\pi^{\mu}$ with $p_{\mu}$ gives
\begin{eqnarray}\label{a.3}
(p\om)\dot {} = p \pi  - |p|^2\om^5, \,\,\,\, (p \pi) \dot {} = -a p\om - |p|^2 \pi^5.
\end{eqnarray}
It is suggestive to look for combinations of variables that decouple the system formed by (\ref{a.2}) and (\ref{a.3}),
for the variables $\om^5$, $\pi^5$, $p \om$ and $p \pi$. We point out that once the equations for $\om^5$, $\pi^5$ are
solved, one can promptly write the solutions for $\om^{\mu}$ and $\pi^{\mu}$ since they obey a closed second order
differential equation
\begin{eqnarray}\label{a.4}
\ddot \om^{\mu} + a \om^{\mu}  = p^{\mu}(\pi^5 + \dot \om^5), \\
\label{a.5} \ddot \pi^{\mu} + a \pi^{\mu}  = p^{\mu}(\dot \pi^5 -a \om^5).
\end{eqnarray}
The system (\ref{a.2})-(\ref{a.3}) may be solved in a sequence of steps. First we define
\begin{eqnarray}\label{a.6}
\pi^5_{\pm} = \pi^5 \pm \frac{p \pi}{|p|}, \\
\label{a.7} \om^5_{\pm} = \om^5 \pm \frac{p \om}{|p|}.
\end{eqnarray}
The second order equations for the variables above are closed for the pairs $(\om^5_{+}, \pi^5_{-})$ and $(\om^5_{-},
\pi^5_{+})$. In fact,
\begin{eqnarray}\label{a.8}
\ddot \om^5_{+} = -2 |p| \pi^5_{-}-(|p|^2+a)\om^5_{+}, \\
\label{a.9} \ddot \pi^5_{-} = -(|p|^2+a)\pi^5_{-}-2a|p|\om^5_{+},
\end{eqnarray}
and
\begin{eqnarray}\label{a.10}
\ddot \om^5_{-} = 2 |p| \pi^5_{+}-(|p|^2+a)\om^5_{-}, \\
\label{a.11} \ddot \pi^5_{+} = -(|p|^2+a)\pi^5_{+}-2a|p|\om^5_{-}.
\end{eqnarray}
The second step for solving the system (\ref{a.2})-(\ref{a.3}) is given by the rotations,
\begin{eqnarray}\label{a.12}
z_{+} = \sqrt{a}\om^5_{+}+ \pi^5_{-}, \\
\label{a.13} z_{-}= \sqrt{a}\om^5_{+} - \pi^5_{-}.
\end{eqnarray}
for (\ref{a.8})-(\ref{a.9}) and
\begin{eqnarray}\label{a.14}
y_{+} = -\sqrt{a}\om^5_{-}+ \pi^5_{+}, \\
\label{a.15} y_{-} = -\sqrt{a}\om^5_{-} - \pi^5_{+},
\end{eqnarray}
for (\ref{a.10})-(\ref{a.11}). Both systems are decoupled. The equations for $z_{\pm}$ and $y_{\pm}$ are
\begin{eqnarray}\label{a.16}
\ddot z_{\pm} = -\om^2_{\pm}z_{\pm}, \\
\label{a.17} \ddot y_{\pm} = -\om^2_{\pm}y_{\pm},
\end{eqnarray}
where $\om_{\pm} = |p| \pm \sqrt{a}$. The general solution for $z_{\pm}$ and $y_{\pm}$ is given by
\begin{eqnarray}\label{a.18}
z_{+} = A \cos (\om_{+}\tau) + B \sin (\om_{+}\tau), \\
\label{a.19}
z_{-} = C \cos (\om_{-}\tau) + D \sin (\om_{-}\tau), \\
\label{a.20}
y_{+} = E \cos (\om_{+}\tau) + F \sin (\om_{+}\tau), \\
\label{a.21} y_{-} = G \cos (\om_{-}\tau) + H \sin (\om_{-}\tau),
\end{eqnarray}
where capital letters are constants of integration. Inverting the expressions (\ref{a.12})-(\ref{a.15}) and
substituting (\ref{a.18})-(\ref{a.21}) provides the solution for $\om^5_{\pm}$, $\pi^5_{\pm}$. In turn, we invert
(\ref{a.6})-(\ref{a.7}) to obtain the solutions
\begin{eqnarray}\label{a.22}
\om^5 = \frac{1}{4 \sqrt{a}}[(A-E)\cos(\om_{+}\tau) + (B-F)\sin(\om_{+}\tau) + \cr
(C-G)\cos(\om_{-}\tau) + (D-H)\sin(\om_{-}\tau)], \\
\label{a.23} \pi^5 = \frac{1}{4}[(A+E)\cos(\om_{+}\tau) + (B+F)\sin(\om_{+}\tau) - \cr
-(C+G)\cos(\om_{-}\tau) - (D+H)\sin(\om_{-}\tau)], \\
\label{a.24} p\om = \frac{|p|}{4 \sqrt{a}}[(A+E)\cos(\om_{+}\tau) + (B+F)\sin(\om_{+}\tau) + \cr
+ (C+G)\cos(\om_{-}\tau) + (D+H)\sin(\om_{-}\tau)], \\
\label{a.25} p \pi = \frac{|p|}{4}[-(A-E)\cos(\om_{+}\tau) + -(B-F)\sin(\om_{+}\tau) + \cr (C-G)\cos(\om_{-}\tau) +
(D-H)\sin(\om_{-}\tau)]
\end{eqnarray}

In order to solve the initial system of equations, we have taken one more time derivative: the variables $y_{\pm}$,
$z_{\pm}$ obey second order differential equations. So, it is necessary to check the consistency of the solutions
(\ref{a.22})-(\ref{a.25}) substituting them in (\ref{a.2}). lhs and rhs must coincide identically, which implies the
restrictions
\begin{eqnarray}\label{a.26}
A+E = B-F, \\
\label{a.27} E-A =B+F.
\end{eqnarray}
It follows that $E=B$, $F=-A$, $G=D$ and $H=-C$. We are now able to solve the equations (\ref{a.4}), (\ref{a.5}) for
$\om^{\mu}$ and $\pi^{\mu}$. Their general solution is given by
\begin{eqnarray}\label{a.28}
\om^{\mu}(\tau) = K^{\mu}\cos (\sqrt{a}\tau) + L^{\mu}\sin (\sqrt{a}\tau) - \cr -\frac{(A+B)}{4|p|\sqrt{a}}p^{\mu} \cos
(\om_{+}\tau) + \frac{(A-B)}{4|p|\sqrt{a}}p^{\mu} \sin (\om_{+}\tau) - \cr - \frac{(C+D)}{4|p|\sqrt{a}}p^{\mu} \cos
(\om_{-}\tau) + \frac{(C-D)}{4|p|\sqrt{a}}p^{\mu} \sin (\om_{-}\tau)
\end{eqnarray}
\begin{eqnarray}\label{a.29}
\pi^{\mu}(\tau) = M^{\mu}\sin (\sqrt{a}\tau) + N^{\mu}\cos (\sqrt{a}\tau) - \cr -\frac{(A+B)}{4|p|\sqrt{a}}p^{\mu} \sin
(\om_{+}\tau) + \frac{(A-B)}{4|p|\sqrt{a}}p^{\mu} \cos (\om_{+}\tau) - \cr - \frac{(C+D)}{4|p|\sqrt{a}}p^{\mu} \sin
(\om_{-}\tau) - \frac{(C-D)}{4|p|\sqrt{a}}p^{\mu} \cos (\om_{-}\tau)
\end{eqnarray}
Substituting the expressions above in the initial system to check its consistency leads to
\begin{eqnarray}\label{a.30}
N^{\mu} =- \sqrt{a}K^{\mu}, \,\, M^{\mu}=\sqrt{a}L^{\mu}, \\
\label{a.31} p_{\mu}K^{\mu} = p_{\mu}L^{\mu} = 0. \qquad \,\,\,\,
\end{eqnarray}
Redefining the constants of integration
\begin{eqnarray}
\frac{A+B}{4} \rightarrow A, \qquad \frac{A-B}{4}
\rightarrow B, \\
\frac{C+D}{4} \rightarrow C, \qquad \frac{C-D}{4} \rightarrow D,
\end{eqnarray}
we are left with the solutions
\begin{eqnarray}\label{a.32}
\om^5(\tau) = \frac{1}{\sqrt{a}}(A \sin(\om_{+}\tau) + B\cos (\om_{+}\tau) +C \sin(\om_{-}\tau)+\cr +D
\cos(\om_{-}\tau)),
\end{eqnarray}
\begin{eqnarray}\label{a.33}
\pi^5(\tau)= A \cos(\om_{+}\tau) - B\sin (\om_{+}\tau) -C \cos(\om_{-}\tau)+\cr +D \sin(\om_{-}\tau),
\end{eqnarray}
\begin{eqnarray}\label{a.34}
\om^{\mu}(\tau) = K^{\mu}\cos (\sqrt{a}\tau) + L^{\mu}\sin (\sqrt{a}\tau) - \cr -\frac{A}{|p|\sqrt{a}}p^{\mu} \cos
(\om_{+}\tau) + \frac{B}{|p|\sqrt{a}}p^{\mu} \sin (\om_{+}\tau) - \cr - \frac{C}{|p|\sqrt{a}}p^{\mu} \cos (\om_{-}\tau)
+ \frac{D}{|p|\sqrt{a}}p^{\mu} \sin (\om_{-}\tau)
\end{eqnarray}
\begin{eqnarray}\label{a.35}
\pi^{\mu}(\tau)= \sqrt{a} L^{\mu}\cos (\sqrt{a}\tau) -\sqrt{a} K^{\mu}\sin (\sqrt{a}\tau) + \cr + \frac{A}{|p|}p^{\mu}
\sin (\om_{+}\tau) + \frac{B}{|p|}p^{\mu} \cos (\om_{+}\tau) - \cr - \frac{C}{|p|}p^{\mu} \sin (\om_{-}\tau) -
\frac{D}{|p|}p^{\mu} \cos (\om_{-}\tau).
\end{eqnarray}
Let us see what restrictions the constraints of the theory impose over the constants of integration. $T_3 = (\pi^A)^2 +
a_3 = 0$ implies
\begin{eqnarray}\label{a.36}
-a_3 = aL^2\cos^2(\sqrt{a}\tau)+ aK^2 \sin^2(\sqrt{a}\tau) - \cr - 2aKL \sin(\sqrt{a}\tau)\cos(\sqrt{a}\tau) -(A^2 +
B^2 +C^2 + D^2) + \cr +2\cos(2 \sqrt{a} \tau)(AC+ BD) + 2 \sin(2 \sqrt{a}\tau) (AD-BC).
\end{eqnarray}
The constraint $T_4 = (\om^A)^2 + a_4 = 0$ leads to (we remind that $a = a_3/a_4$)
\begin{eqnarray}\label{a.37}
-a_3 = aL^2\sin^2(\sqrt{a}\tau)+ aK^2 \cos^2(\sqrt{a}\tau) + \cr + 2aKL \sin(\sqrt{a}\tau)\cos(\sqrt{a}\tau) -(A^2 +
B^2 +C^2 + D^2) - \cr -2\cos(2 \sqrt{a} \tau)(AC+ BD) - 2 \sin(2 \sqrt{a}\tau) (AD-BC).
\end{eqnarray}
Adding the expressions (\ref{a.36}) and (\ref{a.37}), we arrive at
\begin{eqnarray}\label{a.38}
-2a_3 = a(K^2 + L^2) - 2(A^2 + B^2 +C^2 + D^2).
\end{eqnarray}
$T_5 = \om^A \pi^A = 0$ gives
\begin{eqnarray}\label{a.39}
0= \cos(2 \sqrt{a}\tau)\left [KL +\frac{2}{a}(BC-AD) \right ]+ \cr +\sin(2 \sqrt{a}\tau) \left [ \frac{1}{2}(L^2 -
K^2)+ \frac{2}{a}(AC+BD) \right ],
\end{eqnarray}
which must be zero throughout all the time, then
\begin{eqnarray}\label{a.40}
aKL = 2(AD-BC), \\
\label{a.41} a(K^2 - L^2) = 4(AC+BD).
\end{eqnarray}

Now let us construct $J^{AB}$ and $x^{\mu}$. By definition,
$$ J^{5\mu}  =  2(\om^5 \pi^{\mu} - \pi^5 \om^{\mu}). $$
Thus
\begin{eqnarray}\label{a.42}
J^{5\mu}  =  \frac{2(A^2+ B^2 -C^2 - D^2)}{|p|\sqrt{a}}p^{\mu}+ \cr + 2|p| \al^{\mu} \sin(|p|\tau) +2|p| \be^{\mu}
\cos(|p|\tau)
\end{eqnarray}
where
\begin{eqnarray}\label{a.43}
\al^{\mu} =\frac{1}{|p|} [(A+C)L^{\mu}+(B-D) K^{\mu}], \\
\label{a.44} \be^{\mu}= \frac{1}{|p|} [(B+D)L^{\mu}+(C-A) K^{\mu}].
\end{eqnarray}
To write $J^{5\mu}$ we have used the solutions (\ref{a.32})-(\ref{a.35}). There is one more constraint to be taken into
account: $T_2 = p_{\mu}J^{5 \mu}+ mc\hbar = 0$. We have
\begin{eqnarray}\label{a.45}
A^2+ B^2 - C^2 - D^2 = \frac{mc\hbar \sqrt{a}}{2|p|}.
\end{eqnarray}
$J^{\mu \nu}$ can be found in the same way
\begin{eqnarray}\label{a.46}
J^{\mu \nu} &=& 2(\om^{\mu} \pi^{\nu}-\om^{\nu} \pi^{\mu}) \cr &=& \Sigma^{\mu \nu}_1 + \Sigma^{\mu \nu}_2
\sin(|p|\tau)+ \Sigma^{\mu \nu}_3 \cos(|p|\tau),
\end{eqnarray}
where
\begin{eqnarray}\label{a.47}
\Sigma^{\mu \nu}_1 = 2\sqrt{a}(K^{\mu}L^{\nu}-K^{\nu}L^{\mu}),
\qquad \qquad \qquad \quad \\
\label{a.48} \Sigma^{\mu \nu}_2 = \frac{2}{|p|}[(A-C)(K^{\mu}p^{\nu} -
K^{\nu}p^{\mu})-(B+D)(L^{\mu}p^{\nu} - L^{\nu}p^{\mu})], \\
\label{a.49} \Sigma^{\mu \nu}_3 = \frac{2}{|p|}[(A+C)(L^{\mu}p^{\nu} - L^{\nu}p^{\mu})+(B+D)(K^{\mu}p^{\nu} -
K^{\nu}p^{\mu})].
\end{eqnarray}
Since $\dot x^{\mu} = \frac{1}{2}J^{5 \mu}$, one has
\begin{eqnarray}\label{a.50}
x^{\mu}(\tau) = X^{\mu}+ \frac{mc\hbar}{2|p|^2}p^{\mu}\tau- \al^{\mu}\cos(|p|\tau) +\be^{\mu}\sin(|p|\tau).
\end{eqnarray}
We have already used (\ref{a.45}). The physical trajectory $x^i = x^i(t)$ can be found with the same prescription as
before: $x^0$ must be a monotonic function of $\tau$, then we set $K^0 = L^0 = 0$. So, we have (for simplicity, we take
$X^0$)
\begin{eqnarray}\label{a.51}
x^0 = ct = \frac{mc\hbar}{2|p|^2}p^0 \tau \Rightarrow \tau = \frac{2|p|^2}{m\hbar p^0}.
\end{eqnarray}
Hence,
\begin{eqnarray}\label{a.52}
x^i(t) = X^i +c \frac{p^i}{p^0}t - \al^i\cos(\ti \om t)+ \be^i\sin(\ti \om t),
\end{eqnarray}
where $\ti \om = \frac{2|p|^3}{m \hbar p^0}$. We point out that the \textit{Zitterbewegung} takes place in a plane
orthogonal to the vector $p^i$ once the restrictions (\ref{a.31}) are reduced to $\vec p \cdot \vec K = \vec p \cdot
\vec L = 0$, leading to $\vec p \cdot \vec \al = \vec p \cdot \vec \be = 0$. One can also shows that the solution
(\ref{a.42}) for $J^{5 \mu}(\tau)$ obeys the identity
\begin{eqnarray}
(J^{5 \mu}(\tau))^2 + 4[a_3(\om^5(\tau))^2 + a_4 (\pi^5(\tau))^2] \equiv 0,
\end{eqnarray}
that is $J^{5 \mu}$ turns out to be the time-like vector.

In this appendix we have solved the equations of motion for the initial variables $(x^\mu, \om^A, \pi^B)$ and confirmed
the results obtained in Section 6: the variables $J^{AB}$ and $x^{\mu}$ experience \textit{Zitterbewegung} with the
angular frequency $\ti \om = \frac{2|p|^3}{m \hbar p^0}$.

\end{document}